\documentclass[10pt]{article}

\usepackage[english]{babel}

\usepackage[a4paper,top=2cm,bottom=2cm,left=3cm,right=3cm,marginparwidth=1.75cm]{geometry}

\usepackage[backend=biber,style=nature]{biblatex}
\usepackage{amsmath}
\usepackage{amssymb}
\usepackage{graphicx}
\usepackage[colorlinks=true, allcolors=blue]{hyperref}
\usepackage[auth-sc, affil-it]{authblk}
\usepackage{caption}
\usepackage{subcaption}
\usepackage{csquotes}

\usepackage{booktabs}
\usepackage{colortbl}
\usepackage{xcolor}
\usepackage{xfrac}

\usepackage{parskip}
\providecommand{\keywords}[1]
{
  \small	
  \textbf{\textit{Keywords---}} #1
}

\title{A universal crack tip correction algorithm discovered by physical deep symbolic regression}
\author[1,*]{David Melching}
\author[1]{Florian Paysan}
\author[1]{Tobias Strohmann}
\author[1]{Eric Breitbarth}
\affil[1]{German Aerospace Center (DLR), Institute of Materials Research, Linder Hoehe, 51147 Cologne, Germany.}
\affil[*]{Corresponding author: David.Melching@dlr.de}

\date{\today}

\begin{document}
\maketitle

\begin{abstract}
Digital image correlation is a widely used technique in the field of experimental mechanics. In fracture mechanics, determining the precise location of the crack tip is crucial. In this paper, we introduce a universal crack tip detection algorithm based on displacement and strain fields obtained by digital image correlation.
Iterative crack tip correction formulas are discovered by applying deep symbolic regression guided by physical unit constraints to a dataset of simulated cracks under mode I, II and mixed-mode conditions with variable T-stress.
For the training dataset, we fit the Williams series expansion with super-singular terms to the simulated displacement fields at randomly chosen origins around the actual crack tip.
We analyse the discovered formulas and apply the most promising one to digital image correlation data obtained from uniaxial and biaxial fatigue crack growth experiments of AA2024-T3 sheet material. Throughout the experiments, the crack tip positions are reliably detected leading to improved stability of the crack propagation curves.
\end{abstract}



\section{Introduction} \label{sec:intro}

Digital Image Correlation (DIC) is an important tool for investigating crack tip fields in experimental fracture mechanics \cite{Hebert2022,Becker2023,Tong2018}.
Integrated into experiments, this technique can generate large data sets to investigate features such as stress intensity factors \cite{Christopher2013, Roux2009, Becker2012}, crack propagation laws \cite{Mathieu2012,Sanchez2021}, the crack tip plastic zone \cite{Vasco2016, Besel2016} or crack closure \cite{Gonzales2017,Patriarca2018}. Precise information on the coordinates and orientation of the crack tip is essential for quantitative evaluation of each data set. Computer vision techniques are effective for analysing cracks on metal surfaces \cite{Gao2011, Gebauer2022}. However, DIC requires a stochastic speckle pattern, making such techniques inapplicable. The precise identification of the crack path, particularly the crack tip within the strain and displacement field data, is the major challenge. 

In linear-elastic fracture mechanics, the stress and displacement fields around a crack tip can be described in polar coordinates $(r, \theta)$ by the Williams series expansion \cite{Williams1961}:
\begin{align}
\sigma_{ij}(r,\theta) &= \sum_{n} {r^{\frac{n}{2}-1}\ \left( A_n f_{\text{I},ij}(\theta,n) + B_n f_{\text{II},ij}(\theta,n) \right)}, \label{eq:williams_stress} \\
u_{i}(r,\theta) &= \sum_{n} \frac{r^{\frac{n}{2}}}{2\mu} \ \left( A_n g_{\text{I},i}(\theta,n) + B_n g_{\text{II},i}(\theta,n) \right). \label{eq:williams_displ}
\end{align}
The parameters $A_n, B_n \in \mathbb{R}$ are called Williams coefficients and depend on the crack tip loading conditions. The trigonometric functions $f$ and $g$ are known. First order terms are proportional to the stress intensity factors as $K_{\rm I} = \sqrt{2\pi} \cdot A_1$, $K_{\rm II} = -\sqrt{2\pi} \cdot B_1$ and one second order term to the $T$-stress via $T = 4 \cdot A_2$. 

\textit{Crack tip detection.}\\
There are several approaches for detecting cracks in DIC data.
Lopez-Crespo et al. \cite{LopezCrespo2008} applied a Sobel line detection algorithm to the vertical displacement field to locate the crack tip. The crack tip position can also be determined by using a least-squares fitting of the truncated Williams' expansion and including the crack tip coordinates in the feature set \cite{Yoneyama2006}. Yang et al. \cite{Yang2021} followed a similar approach, but excluded the plastic zone. By incorporating negative Williams series terms, also known as super-singular terms, Réthoré \cite{Rethore2015} derived the iterative correction formula \ref{eq:rethore} to find the position of the crack tip under pure mode I loading. 

\begin{equation} \label{eq:rethore}
    d_{x}=-2 \cdot \frac{A_{-1}}{A_{1}}
\end{equation}

Following this formula, after some iterations $A_{-1}$ tends to zero. This approach is constrained to a given crack plane and an initial estimate for the crack tip is required.
Baldi and Santucci \cite{Baldi2022} built upon Réthoré's \cite{Rethore2015} approach and discovered that $A_{-1}$ and $B_{-1}$ exhibit linear behaviour around the crack tip and use two planes to detect it at $A_{-1}=B_{-1}=0$. 
Cinar et al. \cite{Cinar2017} proposed a new phase congruency based algorithm to automatically segment cracks and extract their quantifying parameters such as crack path, length and opening displacement. 
To address the problem of noisy displacement data, Gupta et al. \cite{Gupta2023} introduced a separability approach that uses the multiplicative separability of the asymptotic stress field. 
Zanganeh et al. \cite{Zanganeh2012} performed a comparative study of different optimization methods to identify the crack tip in displacement data by fitting of the Williams series. 
Further work by Bonniot et al.\cite{Bonniot2019} coupled a grid search algorithm \cite{Harilal2015} for an initial guess on a coarse grid evaluation with the pattern search method for an iterative detection of the crack tip location. 
Shuai et al. \cite{Shuai2022} first determine the plastic zone. Then two symmetrical lines are defined on both sides of the crack. The theoretical displacement difference between the two lines is analysed to estimate the crack tip position. 
Similarly, Broggi et al.\cite{Broggi2023} used a technique to determine the effective length of a crack by analysing crack opening displacement (COD) profiles obtained from displacement fields. 
Gehri et al. \cite{Gehri2020} introduced a method to detect multiple cracks using the DIC principal tensile strain field. The method is capable of detecting complex crack paths including branching and bifurcation but is affected by the spatial resolution of DIC as well as inherent scatter, noise and artefacts. Panwitt et al. \cite{Panwitt2022} extended this method to determine the position of the tips by using the crack opening.
To overcome the general problem of artefacts and scatter in DIC data, Strohmann et al. \cite{Strohmann2021} developed a machine learning model based on a U-Net architecture, which was trained to accurately detect both the crack path and crack tip using full-field DIC displacement data. Melching et al. \cite{Melching2022} combined this U-Net model at its deepest layer with a fully connected neural network and demonstrated the use of explainable AI for selecting models which generalize well.

\textit{Symbolic regression.}\\
Although effective, machine learning models can lack transparency, which is problematic for certain applications or when applied to new categories of data. Consequently, efficient analytical formulas and algorithms are preferred \cite{Camps-Valls2023}. Symbolic regression is a machine learning methodology that seeks to automatically identify formulas representing correlations in a given data set \cite{Angelis2023}. By applying the Buckingham-Pi theorem, these methods can even consider physical units to derive analytical physical formulas by exploring the space of available functional forms \cite{Tenachi2023}. 

\textit{Our work.}\\
In this work, we use symbolic regression to discover crack tip correction formulas based on Williams series coefficients.
Our preliminary numerical studies have shown that the energy landscape gradients near the crack tip vanish, making it difficult to determine the crack tip position accurately. To address the stated limitations, Physical Symbolic Optimization ($\Phi$-SO) \cite{Tenachi2023} is utilized to identify formulas for efficient crack tip correction in the $x$- and $y$- directions for mode I, II, and mixed-mode loading scenarios. Initially, we create a finite element (FE) model and run linear-elastic simulations with exact knowledge of the crack tip position under several external loading scenarios. Then, we calculate the Williams coefficients at various randomly chosen positions around the actual crack tip using the over-deterministic fitting method \cite{Ayatollahi2011} implemented in \textsc{CrackPy} \cite{strohmann_2022_7319653}. The resulting data set is used to learn correction formulas for $d_x$ and $d_y$ using the physical deep symbolic regression tool $\Phi$-SO. Among the discovered Pareto formulas, we select the one that exhibits wide applicability among different load cases and optimal convergence properties. Finally, we apply an this formula to multiscale DIC data obtained from uniaxial and biaxial fatigue crack propagation experiments. To estimate the crack tip initially, we introduce a line interception method based on the characteristic displacement gradients near the crack path.

\section{Methodology} \label{sec:method}

\subsection{Training data}\label{sec:method_fem}
To generate the training data set, we parameterized a 2D FE model using \textsc{pyansys} \cite{alexander_kaszynski_2020_4009467}. Figure \ref{fig:approach} illustrates the single edge notched model's geometry and boundary conditions. The sheet has a quadratic shape of $w \times h = 100\times100\,\mathrm{mm^2}$ with a crack length $a = 50\,\mathrm{mm}$, i.e. $a/w=0.5$, ensuring symmetry within the model. We used a structured mapped mesh of rectangular 4-node elements (PLANE182) with an element edge length of $0.2\,\mathrm{mm}$ and plane stress formulation. This is a suitable element size for evaluating the Williams coefficients \cite{Melching2023}. Our model consists of 250000 elements and 251251 nodes. We used a linear elastic material formulation with a Young's modulus $E=72\,\mathrm{GPa}$ and a Poisson's ratio $\nu_{xy}=0.33$, which is typical for aluminium alloys. The model was fixed in the centre of the coordinate system using displacement boundary conditions. This model allows the application of defined boundary conditions so that the stress intensity factors $K_{\mathrm{I}}$ and $K_{\mathrm{II}}$ are directly related to $\sigma_{yy}$ and $\sigma_{xy}$, respectively. However, $\sigma_{xx}$ and $\sigma_{yy}$ affect the $T$-stress. 
According to the given boundary conditions, we can linearly approximate $K_{\mathrm{I}} \approx 1.184 \cdot \sigma_{yy} \, \sqrt{\mathrm{m}} $, $K_{\mathrm{II}} \approx 0.541 \cdot \sigma_{xy} \, \sqrt{\mathrm{m}} $, and  $T \approx \sigma_{xx} + 0.658 \cdot \sigma_{yy}$.
The results were exported as tabular data containing the nodal coordinates, displacements and the total strains.

\subsection{Data generation} \label{sec:method_data}
Using the model in Section \ref{sec:method_fem}, we generate several FE simulations by varying the boundary conditions $\sigma_{xx}$, $\sigma_{yy}$, and $\sigma_{xy}$. Since we use a linear-elastic model, only few combinations of boundary conditions are necessary to achieve a sufficiently large variation. The choice of boundary conditions is as follows:

\begin{itemize}
 \item $\sigma_{xx} \in \{-10.0, 0.0, 10.0\}$
 \item $\sigma_{yy} \in \{0.0, 10.0, 20.0\}$
 \item $\sigma_{xy} \in \{-10.0, 0.0, 10.0\}$
\end{itemize}

Excluding the trivial case $\sigma_{xx}=\sigma_{yy}=\sigma_{yy} = 0$, this  set of boundary conditions results in 26 FE simulations encompassing pure mode I, mode II, and mixed-mode scenarios.

\begin{figure}[ht]
    \centering
    \includegraphics[width=\linewidth]{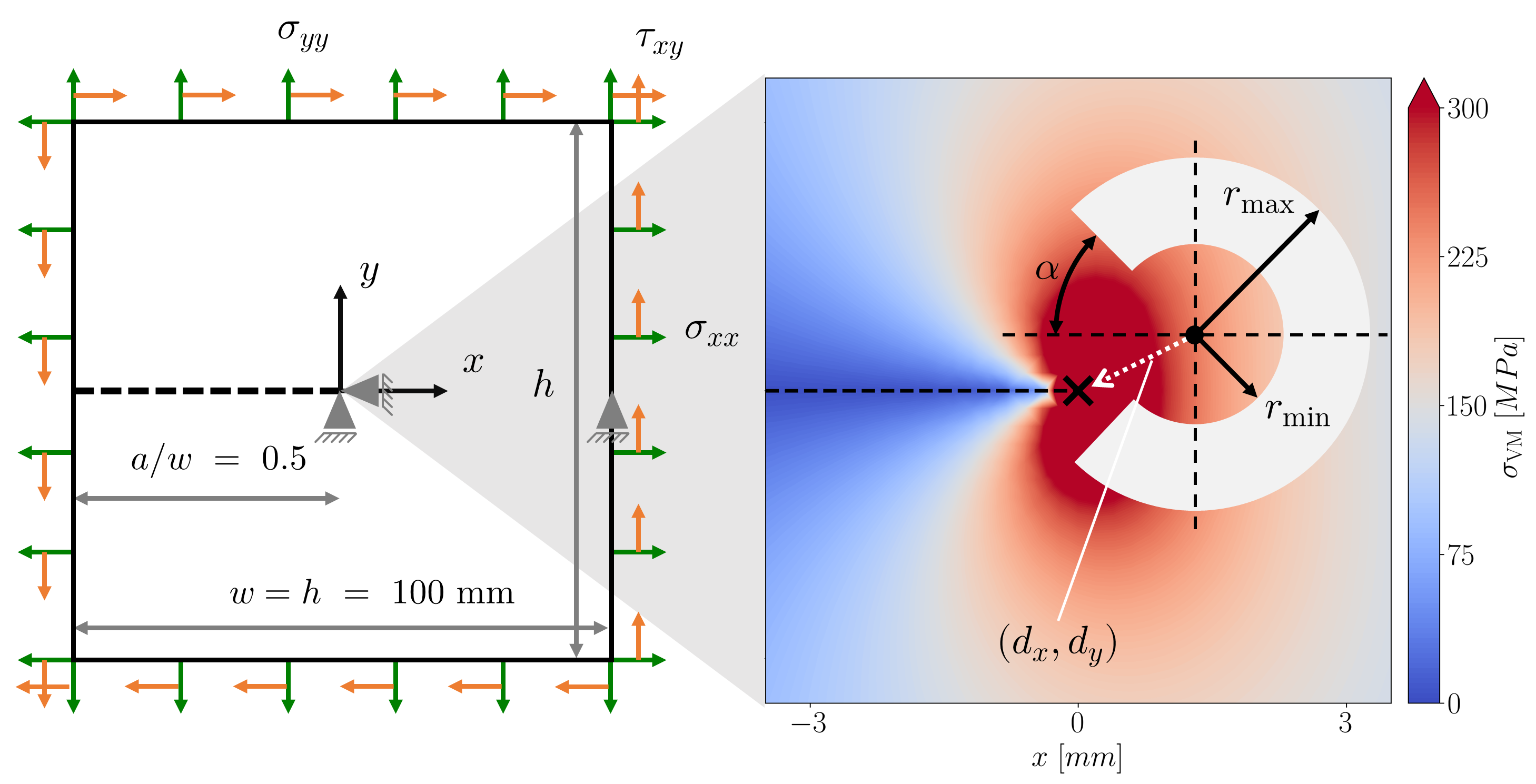}
    \caption{Left: Finite element model with crack and boundary conditions. Right: Random sample point for training of symbolic regression together with the corresponding target correction vector $(d_x,d_y)$ (in white) and angular Williams fitting domain (in gray). The background shows the von Mises eqv. stress for the boundary conditions $\sigma_{xx}=-10 \, \mathrm{MPa}$, $\sigma_{yy} = 20 \, \mathrm{MPa}$, $\tau_{xy} = 0 \, \mathrm{MPa}$.}
    \label{fig:approach}
\end{figure}

The left-hand side of Figure \ref{fig:approach} shows the FE model together with the subdomain of the possible crack tip estimates. The right-hand side illustrates the correction scheme. At a randomly chosen point of the subdomain, we fit the Williams series expansion to the displacement field using an angular fitting domain.

For every FE simulation, we randomly choose 1000 points from a $6 \times 6 \,mm^2$ subdomain around the actual crack tip and fit the Williams series expansion to the FE data at these points. This subdomain can be interpreted as region of possible crack tip estimation. We consider the size of this domain to be appropriate, as we can also estimate the crack tip with comparable accuracy in an experiment. Consequently, the whole data set consists of $26 \times 1000 = 26000$ samples usable for training the symbolic regression model. 

Heading for iterative crack tip correction, we effectively treat these random points as our current estimate for the crack tip position. It should be noted that the Williams series coefficients can be calculated even if the origin, i.e. the estimated crack tip position, is not aligned with the actual crack tip position. To calculate the Williams series coefficients at each of these random points, we use the over-deterministic fitting method \cite{Ayatollahi2011} implemented in \textsc{CrackPy} \cite{strohmann_2022_7319653} version 1.2.0 (\url{https://github.com/dlr-wf/crackpy/}).

Therefore, we use an angular fitting domain with a missing segment of $\alpha = 45\, ^\circ$ around the crack path, an internal radius of $r_{\rm min} = 5 \, \mathrm{mm}$ external radius $r_{\rm max} = 10 \, \mathrm{mm}$. The interpolation tick size for the fitting domain is set to $0.25 \, \mathrm{mm}$ and the terms $-3 \leqslant n \leqslant 7$ of the Williams series expansion were chosen as free parameters for $A_n$ and $B_n$, resulting in 22 parameters for each sample point. Additionally, for each sample point, we know the ground truth correction vector components $d_x$, $d_y$.

\subsection{Physical symbolic regression}\label{sec:method_SR}
Our aim is to find an analytical formula for crack tip correction depending on the Williams coefficients data introduced in Section \ref{sec:method_data}. This approach should effectively generalize the iterative correction Formula \ref{eq:rethore} by Rethoré \cite{Rethore2015} to correct also in $y$-direction and be applicable to more general load cases. To achieve this, we employ the symbolic regression framework \textit{Physical Symbolic Optimization} ($\Phi$-SO) as introduced by Tenachi et al. \cite{Tenachi2023}. This framework is based on deep symbolic regression \cite{Petersen2021} and is implemented in Python (\url{https://github.com/WassimTenachi/PhySO}). It can be used to identify analytical formulas from (noisy) data taking physical unit constraints into account during the equation generation process. This approach drastically reduces the (exponentially large) search space. To this end, we first define the base units of the Williams coefficients. For every $n$, the units of the Williams coefficients $A_n$ and $B_n$ are given by \cite{Karihaloo2003}:
\begin{equation} \label{eq:Williams_unit}
    \text{MPa} \cdot \text{mm}^{1-n/2} = \frac{\text{N}}{\text{mm}^{2}} \cdot \text{mm}^{1-n/2} = \text{N} \cdot \text{mm}^{-1-n/2}
\end{equation}

We conduct six distinct training runs to derive formulas for $d_x$ and $d_y$ for mode I, mode II and mixed-mode loading conditions. The aim is to find a formula for the correction vector that leads from a random starting point to the known crack tip position. We include standard mathematical operators, while excluding trigonometric functions since no periodicity is expected. We allow the candidate functions to contain learnable free constants ($k, m, n$) with fixed units to cover situations, where the problem has unknown physical scales, as well as the fixed constants $1, 2,$ and $4$. $\Phi$-SO generates and evaluates batches of symbolic functions with a recurrent neural network, improving them over time by reinforcing high-reward behaviors. This process helps the network to learn parameters that produce effective symbolic functions. The result is a Pareto front showing the most accurate expression based on the root mean squared error (RMSE) between crack tip correction predictions and targets for each level of formula complexity. The hyperparameters are equal for all training runs and summarized in Table \ref{table:settings_physo}. 

\begin{table}[htbp!]
\centering
\begin{tabular}{|c c|} 
\hline
\ Parameters & Value  \\ [0.5ex]
\hline
Operators & mul, add, sub, div, abs, inv, n2, neg, exp, log \\
Fixed constants & 1, 2, 4 \\
Free constants &  $k$ [1], $m$ [mm], $n$ [N] \\
Reward function & SquashedNRMSE \\
Input & $A_n$, $B_n$ with $-3 \leqslant n \leqslant 7$ \\
Target & $d_x$, $d_y$ \\
Batch size & 1000 \\
Epochs & 1000 \\
\hline
\end{tabular}
\caption{Training settings for $\Phi$-SO}
\label{table:settings_physo}
\end{table}

\section{Results and discussion} \label{sec:results}
We use the Williams coefficients data set described in Section \ref{sec:method_data} to train crack tip correction models using Physical Symbolic Optimization (see Section \ref{sec:method_SR}. Each training run is performed with the hyperparameters given in Table \ref{table:settings_physo} but the data varied as follows:

\begin{itemize}
    \item Mode I: All boundary conditions except if $\sigma_{xy} \neq 0$ or $\sigma_{yy} = 0$
    \item Mode II: All boundary conditions except if $\sigma_{xy} = 0$ or $\sigma_{yy} \neq 0$
    \item Mixed-mode: All boundary conditions
\end{itemize}

\begin{figure}[htbp!]
    \centering
    \includegraphics[width=0.49\textwidth]{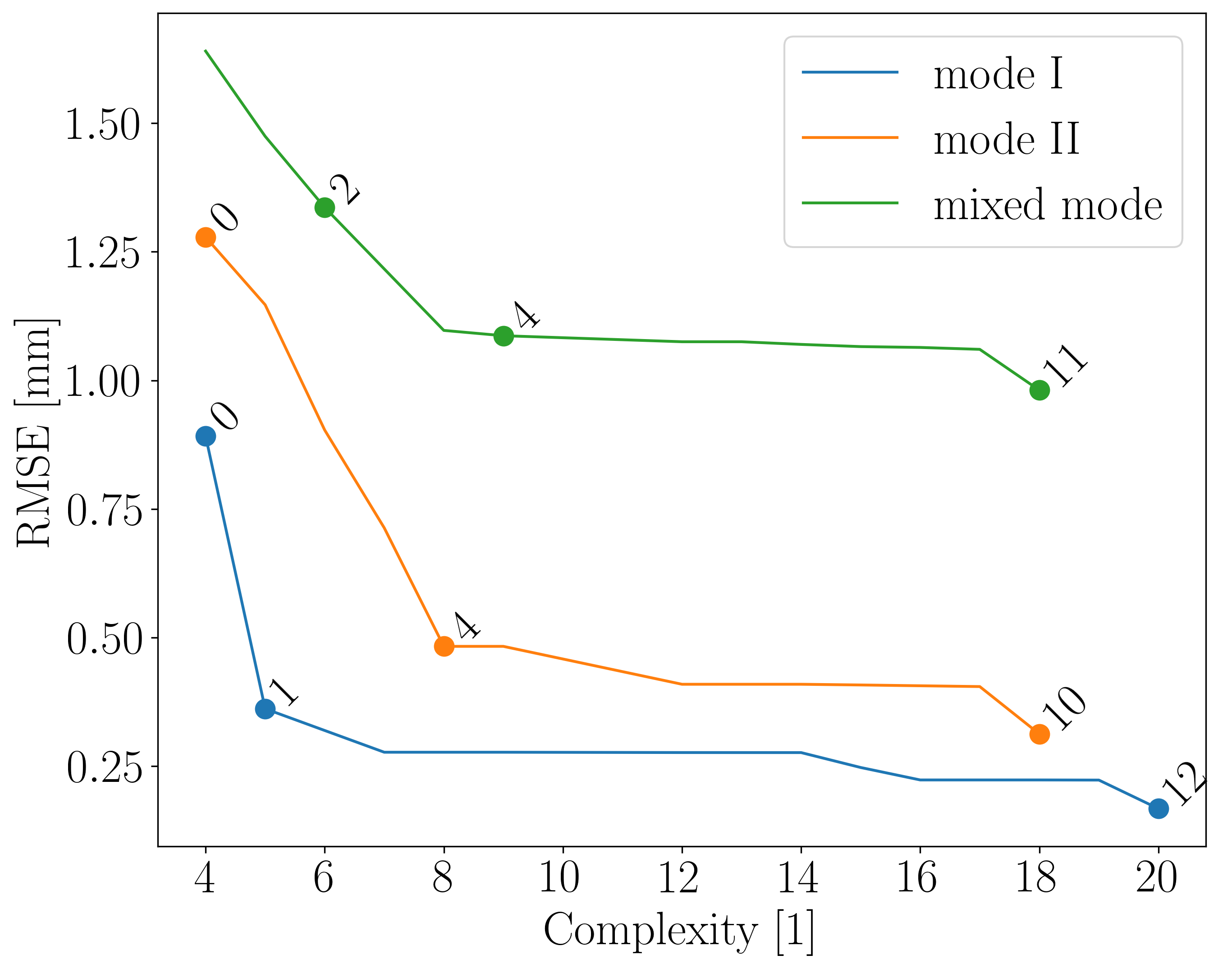}
    \includegraphics[width=0.49\textwidth]{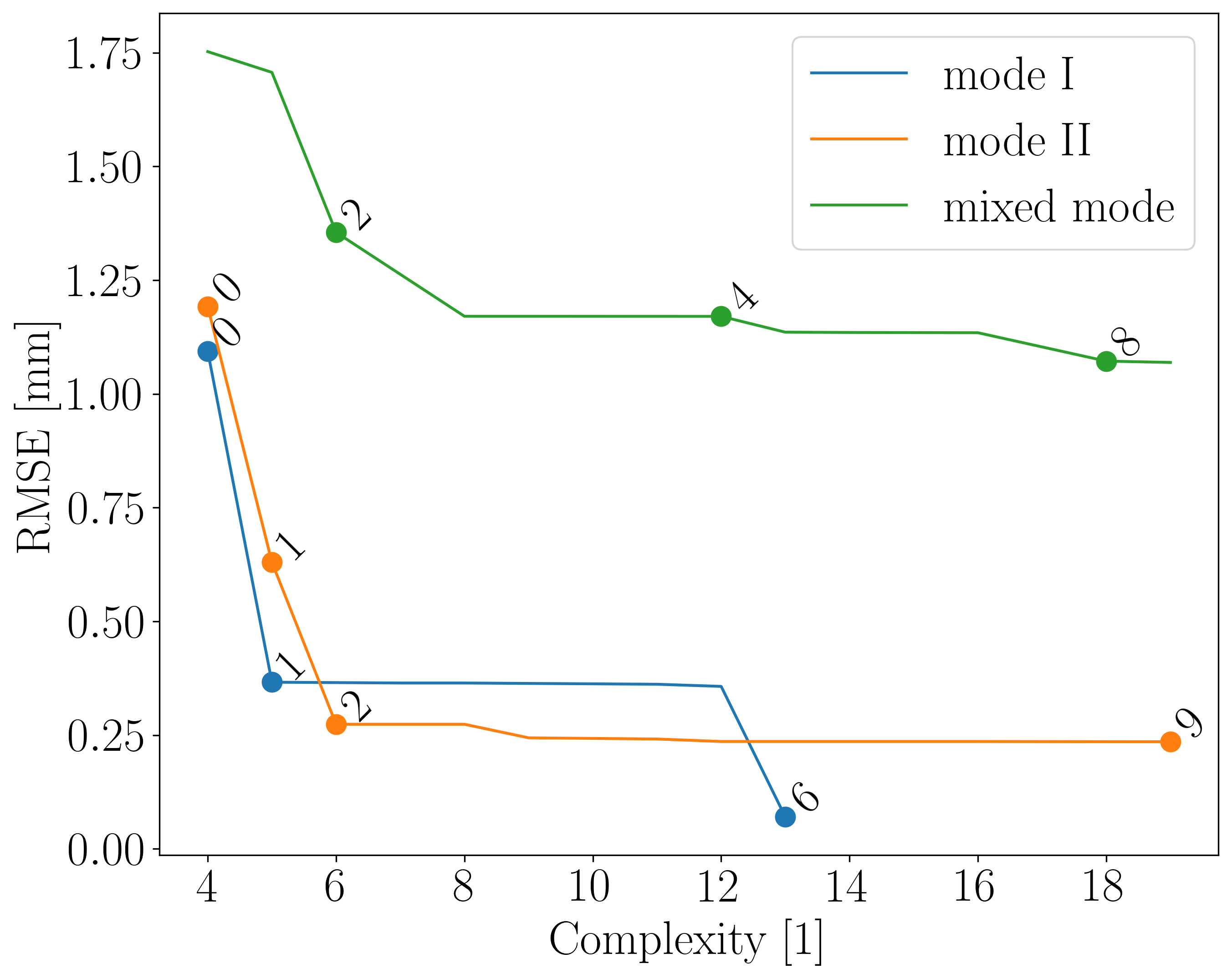}
    \caption{Pareto fronts for different training sets. Left: $x$-correction. Right: $y$-correction. The numbers correspond to the \#-column in Tables \ref{table:eqs_pareto_I}-\ref{table:eqs_pareto_mixed}.}
    \label{fig:pareto_fronts}
\end{figure}

Since $\Phi$-SO only allows to train scalar functions, the correction formulas in $x$- and $y$-direction are learned separately. For each run, we receive a Pareto front of crack tip correction formulas with minimal root mean-squared error (RMSE) loss given a mathematical complexity. Figure \ref{fig:pareto_fronts} shows these Pareto fronts for the $x$- and $y$-correction.

Because of Pareto optimality, the RMSE decreases with increasing formula complexity. For mode I and mode II the RMSE decreases until a saturation is reached around a complexity of 6-8 with an RMSE of $0.25$-$0.5\,\mathrm{mm}$ for both, the $x$- and $y$-correction, respectively. However, for the mixed-mode case, higher losses remain. This indicates that a simple closed formula, which works for arbitrary mixed-mode load cases might not exist. Representative formulas are shown in Tables \ref{table:eqs_pareto_I}-\ref{table:eqs_pareto_mixed}. Since the Pareto front contains many similar, repetitive formulas, we only display a selection here. 

\begin{table}[htbp!]
\centering
 \begin{tabular}{|c|c c c c c|} 
\hline
& \# & Complexity & Reward & RMSE & Equation \\ [0.5ex]
\hline
& 0 & 4 &  0.66272 &  0.89169 & $- \frac{A_{-1}}{A_{1}}$\\
$d_x$ & 1 & 5 &  0.82885 &  0.36178 & $- \frac{1.90537 A_{-1}}{A_{1}}$\\
& 12 & 20 &  0.91267 &  0.16766 & $- 0.00193 \mathrm{mm} - \frac{0.00276 A_{-2} \mathrm{mm}}{\mathrm{N}} - \frac{2.08617 A_{-1}}{A_{1}}$\\
\hline
& 0 & 4 &  0.61570 &  1.09381 & $- \frac{B_{-1}}{A_{1}}$\\
$d_y$ &1 & 5 &  0.82705 &  0.36646 & $- \frac{2.50796 B_{-1}}{A_{1}}$\\
& 6 & 13 &  0.96146 &  0.07025 & $\frac{9.78182 B_{1} \mathrm{mm} - 1.66341 B_{-1}}{A_{1}}$\\
\hline
\end{tabular}
\caption{Mode I correction formulas on Pareto front}
\label{table:eqs_pareto_I}
\end{table}

\begin{table}[htbp!]
\centering
\begin{tabular}{|c|c c c c c|} 
\hline
& \# & Complexity & Reward & RMSE & Equation \\ [0.5ex]
\hline
& 0 & 4 &  0.57814 &  1.27848 & $- \frac{B_{-1}}{B_{1}}$\\
$d_x$ & 4 & 8 &  0.78389 &  0.48304 & $17.06448 \mathrm{mm} - \frac{0.12296 B_{1}^{2}}{B_{2}^{2}}$\\
& 10 & 18 &  0.84853 &  0.31276 & $- \frac{2.43992 B_{-1}}{B_{1}} + \frac{0.27118 B_{-2}}{B_{2} \mathrm{mm}}$\\
\hline
& 0 & 4 &  0.59518 &  1.19191 & $- \frac{A_{-1}}{B_{1}}$\\
$d_y$ & 1 & 5 &  0.73554 &  0.63006 & $- \frac{2.62186 A_{-1}}{B_{1}}$\\
& 2 & 6 &  0.86479 &  0.27400 & $- \frac{2.29023 A_{-3}}{B_{1} \mathrm{mm}}$\\
& 9 & 19 &  0.88148 &  0.23561 & $- \frac{2.19136 A_{7} A_{-3} \mathrm{mm}^{2}}{B_{1}^{2}}$\\
\hline
\end{tabular}
\caption{Mode II correction formulas on Pareto front}
\label{table:eqs_pareto_II}
\end{table}

\begin{table}[htbp!]
\centering
\begin{tabular}{|c|c c c c c|} 
\hline
& \# & Complexity & Reward & RMSE & Equation \\ [0.5ex]
\hline
& 2 & 6 &  0.56735 &  1.33602 & $- \frac{A_{-1}}{A_{1} + B_{1}}$\\
$d_x$ & 4 & 9 &  0.61716 &  1.08681 & $- \frac{0.00735 A_{-1}}{A_{1}}$\\
& 11 & 18 &  0.64105 &  0.98101 & $\frac{0.0429 A_{1} A_{-1}}{A_{3} \cdot \left(2.88427 A_{1} \mathrm{mm} - A_{-1}\right)}$\\
\hline
& 2 & 6 &  0.56391 &  1.35510 & $- \frac{B_{-1}}{A_{1} + B_{1}}$\\
$d_y$ & 4 & 12 &  0.59954 &  1.17044 & $- \frac{0.0001 B_{-1}}{A_{1}}$\\
& 8 & 18 &  0.62042 &  1.07207 & $\frac{0.02276 B_{-1}}{A_{3} \mathrm{mm}}$\\
\hline
\end{tabular}
\caption{Mixed-mode correction formulas on Pareto front}
\label{table:eqs_pareto_mixed}
\end{table}

For mode I, we find formulas of the form
\begin{equation}\label{eq:symreg_mode_I}
    d_x = - c_x^{\rm I} \frac{A_{-1}}{A_1}, \qquad d_y = - c_y^{\rm I} \frac{B_{-1}}{A_1},
\end{equation}
with constants $c_x^{\rm I}, c_y^{\rm I} > 0$. With Formula \#1 in Table \ref{table:eqs_pareto_I}, we effectively rediscover the iterative crack tip correction algorithm proposed by Rethoré \cite{Rethore2015} with $c_x=2$ (cf. Equation \eqref{eq:rethore}). In addition, Formula \#1 in Table \ref{table:eqs_pareto_I} suggests that for the mode I case a constant $c_y = 5/2$ works best for an iterative crack tip correction in $y$-direction. 

For mode II, the symbolic regression model discovers formulas with low complexity of the form
\begin{equation}\label{eq:symreg_mode_II}
    d_x = - c_x^{\rm II} \frac{B_{-1}}{B_1}, \qquad d_y = - c_y^{\rm II} \frac{A_{-1}}{B_1},
\end{equation}
with constants $c_x^{\rm II}, c_y^{\rm II} > 0$, whereas the more complex formulas with a smaller error contain higher order terms as well. Although getting closer to the crack tip in only one step, these more complex expressions often only work for the specific load case or fail when applied iteratively (see Appendix).

For mixed-mode, we discover the formulas
\begin{equation}\label{eq:symreg_mixed_mode}
    d_x = - \frac{A_{-1}}{A_1+B_1} \qquad \text{and} \qquad d_y = - \frac{B_{-1}}{A_1+B_1}
\end{equation}
which do not contain unit constants and are very similar to the mode I and mode II formulas above. The denominator $A_1+B_1$ solves the problem of vanishing $A_1$ or $B_1$ and the corresponding division by zero in Equations \ref{eq:symreg_mode_I} and \ref{eq:symreg_mode_II} for pure mode I or mode II, respectively. Nevertheless, we will see that these equations still do not work in all scenarios and distinction between mode-I-dominated and mode-II-dominated load cases is necessary when applying the crack tip correction (see Section \ref{sec:application}). We remark that for pure mode I loadings, $B_1=0$ and thus Formula \eqref{eq:symreg_mixed_mode} equals Formula \eqref{eq:symreg_mode_I}. The formulas revealed highlight the advantages of symbolic regression, as only the most important variables are determined. In particular, higher order terms were omitted for plausibility reasons, as their determination becomes increasingly unstable at higher orders with the over-deterministic method. From a set of 22 potential parameters, $A_{1}$, $A_{-1}$, $B_{1}$ and $B_{-1}$ proved to be the most important.

\subsection{Correction vector fields} \label{sec:results_vector_fields}
For a better understanding of the highlighted formulas, this section will concentrate on the correction vectors at various estimated crack tip positions. All constants $c_x^{\rm I}$, $c_y^{\rm I}$,  $c_x^{\rm II}$, $c_y^{\rm II}$ are set to 1, as they are only identified as scale factors and do not affect the general behaviour significantly. Tracing the vector field from a random starting point should iteratively lead to the actual position of the crack tip at $x=y=0$. According to the chosen training data, we select a starting point range, interpretable as \textit{initial crack tip estimation}, of $-3 \, \mathrm{mm} < x,y < 3 \, \mathrm{mm}$.

\begin{figure}[!htb]
    \centering
    \includegraphics[width=0.49\textwidth]{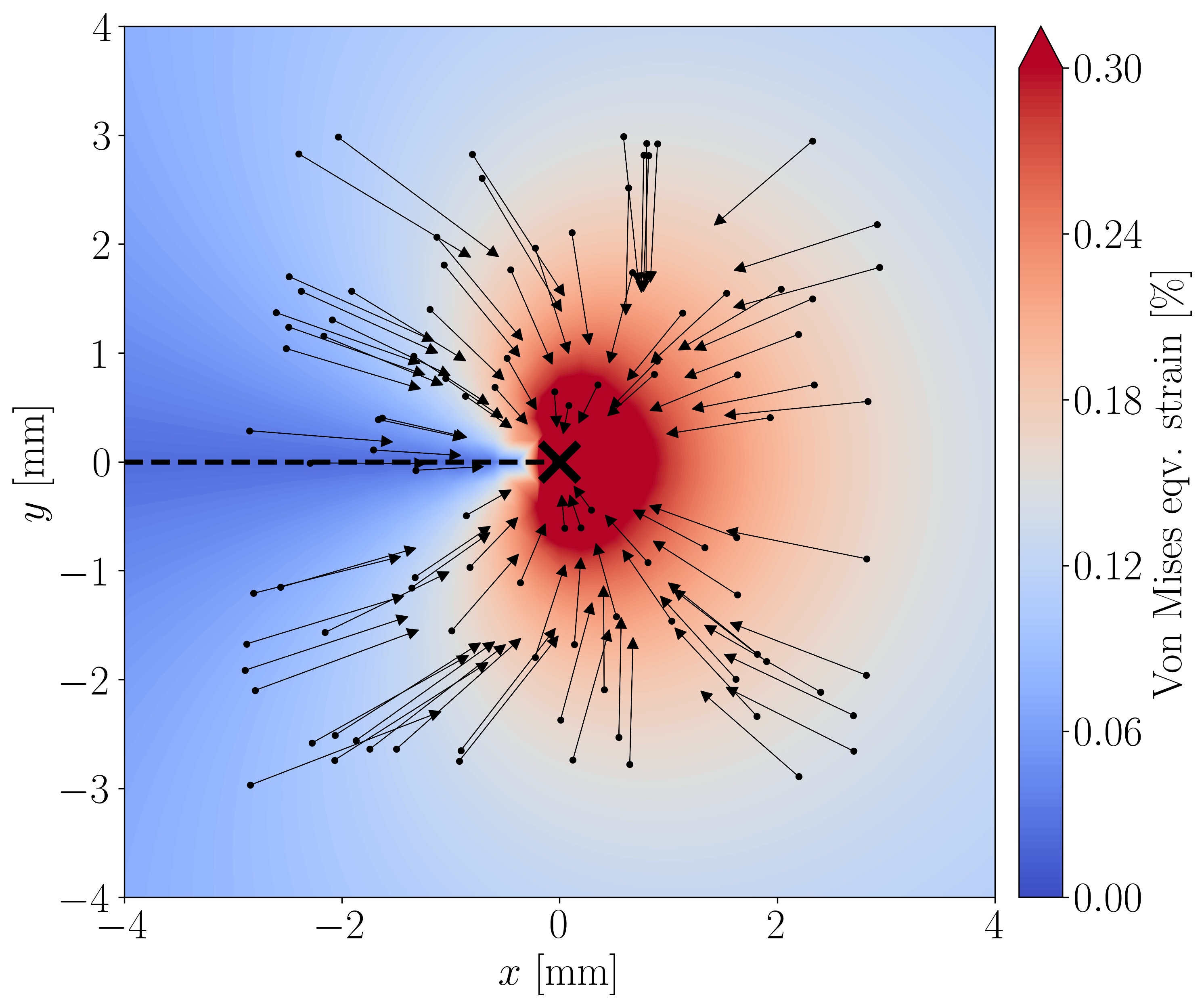}
    \includegraphics[width=0.49\textwidth]{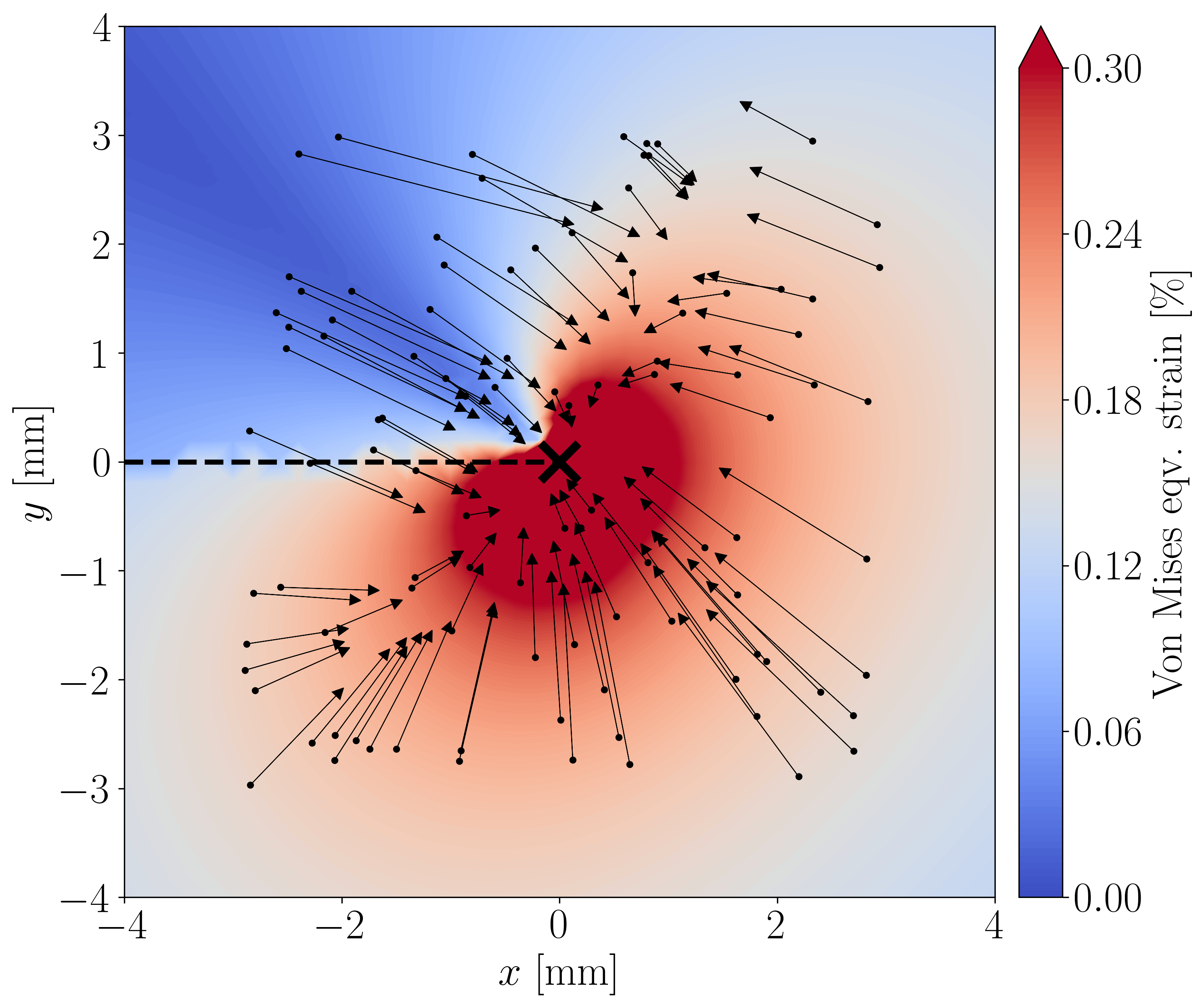}
    \caption{Correction vectors using the mode I formula \eqref{eq:symreg_mode_I} with $c_x^{\rm I} = c_y^{\rm I} = 1$ for the mode I load case $\sigma_{xx} = \sigma_{yy} = 10, \sigma_{xy} = 0$ [MPa] (left) and the mixed-mode load case $\sigma_{xx} = \sigma_{yy} = \sigma_{xy} = 10$ [MPa] (right).}
    \label{fig:vector_plot_mode_I}
\end{figure}

Figure \ref{fig:vector_plot_mode_I} illustrates that all correction vectors point inwards in the direction of the crack tip showing that Formula \eqref{eq:symreg_mode_I} works well under mode I and mixed-mode loadings. However, the mode I formulas do not work for the pure mode II case (see Appendix). 

On the other hand, Figure \ref{fig:vector_plot_mode_II} shows the vector fields for a pure mode II and a mixed-mode load case using the mode II correction formulas \eqref{eq:symreg_mode_II}. We see that the correction vector field for the pure mode II case forms a field driving the iterative correction towards the actual crack tip, while this is not the case for the mixed-mode loading. 

\begin{figure}[!htb]
    \centering
    \includegraphics[width=0.49\textwidth]{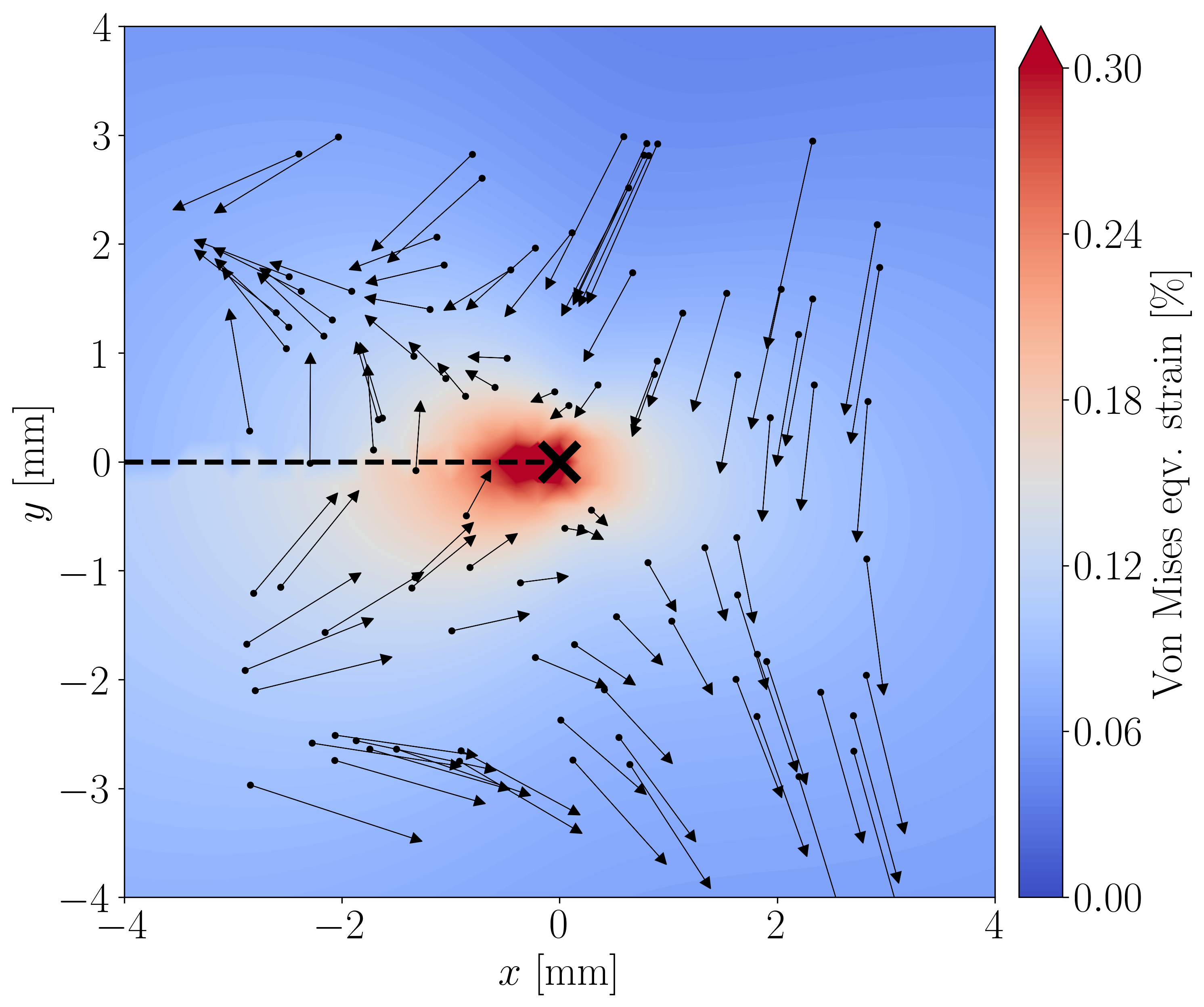}
    \includegraphics[width=0.49\textwidth]{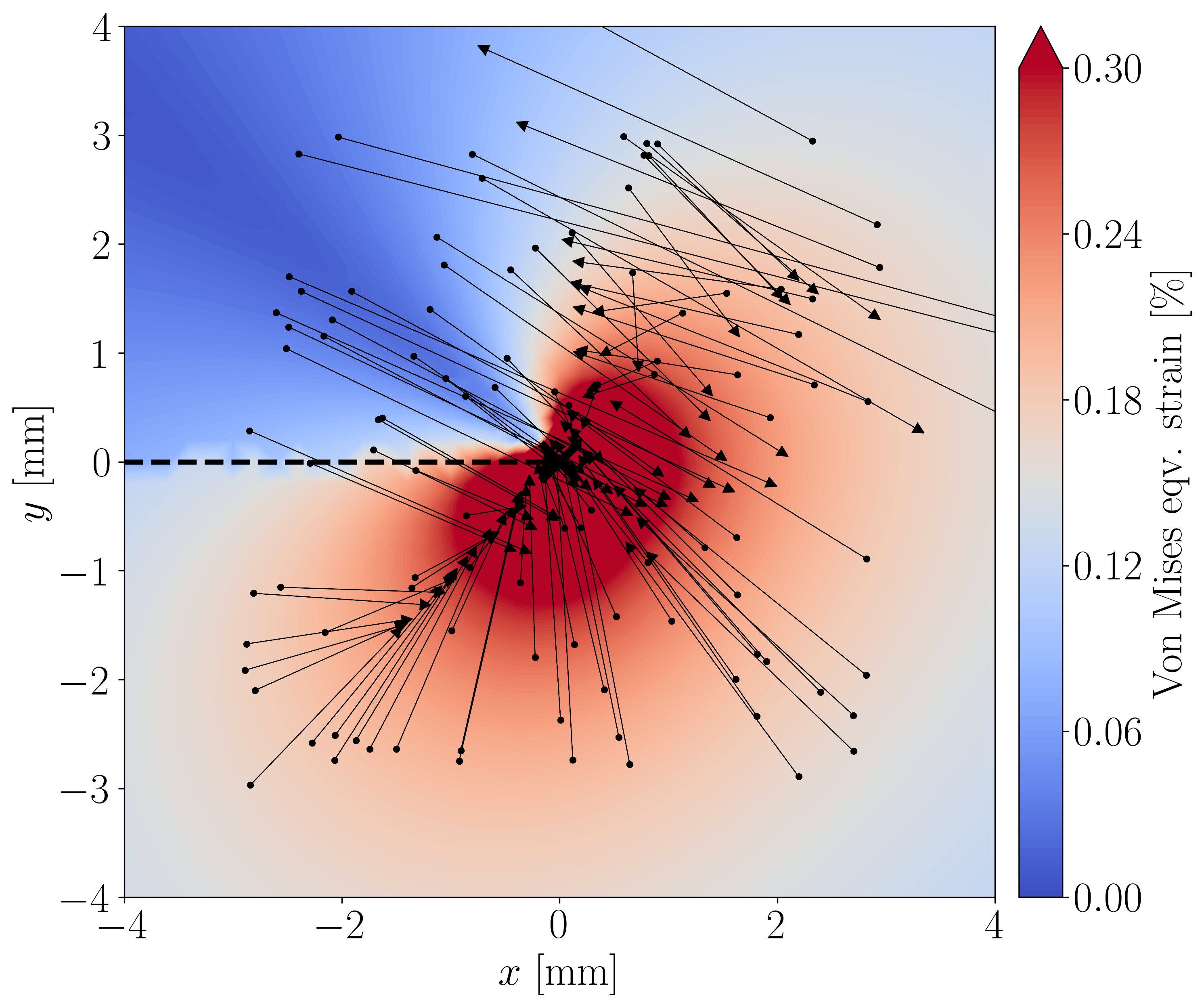}
    \caption{Correction vectors using the mixed-mode formulas \eqref{eq:symreg_mixed_mode} for the mode II load case $\sigma_{xx} = \sigma_{xy} = 10, \sigma_{yy} = 0$ [MPa] (left) and the mixed-mode load case $\sigma_{xx} = \sigma_{yy} = \sigma_{xy} = 10$ [MPa] (right).}
    \label{fig:vector_plot_mixed_mode}
\end{figure}

In Figure \ref{fig:vector_plot_mixed_mode}, we plotted the vector fields for the same load examples using Formula \eqref{eq:symreg_mixed_mode} discovered by the symbolic regression model trained on all load cases (mode I, mode II, mixed-mode). For pure mode I, Formula \eqref{eq:symreg_mixed_mode} becomes Formula \eqref{eq:symreg_mode_I} with $c_x^{\rm I}=c_y^{\rm I}=1$ and the plot therefore corresponds to the left-hand side of Figure \ref{fig:vector_plot_mode_I}. For the pure mode II example on the left-hand side of Figure \ref{fig:vector_plot_mixed_mode}, we observe that the vectors point in the direction of the crack tip only close to the diagonal ($x \approx y$) but away from the crack tip close to the anti-diagonal ($x \approx -y$) suggesting that for initial crack tip estimates in this area the correction will lead away from the crack tip and diverge. For the mixed-mode example on the right-hand side, some vectors point towards the crack tip, however, especially for $x>0, y>0$, the correction vectors point away from the real crack tip. This indicates that the use of this formula for iterative correction might cause convergence issues as well.

We conclude that Formula \eqref{eq:symreg_mode_I} works under mode I as well as mode-I-dominated mixed-mode scenarios, whereas Formula \eqref{eq:symreg_mode_II} only works for pure mode II loadings. Although Formula \eqref{eq:symreg_mixed_mode} looks promising, considering the correction vector fields above, we expect that it does not lead to improved results compared to Formula \eqref{eq:symreg_mode_I}. 

The correction vector fields for additional pairs of formulas discovered by $\Phi$-SO in Tables \ref{table:eqs_pareto_I}-\ref{table:eqs_pareto_mixed}, can be found in the Appendix.
\newpage

\begin{figure}[!htb]
    \centering
    \includegraphics[width=0.49\textwidth]{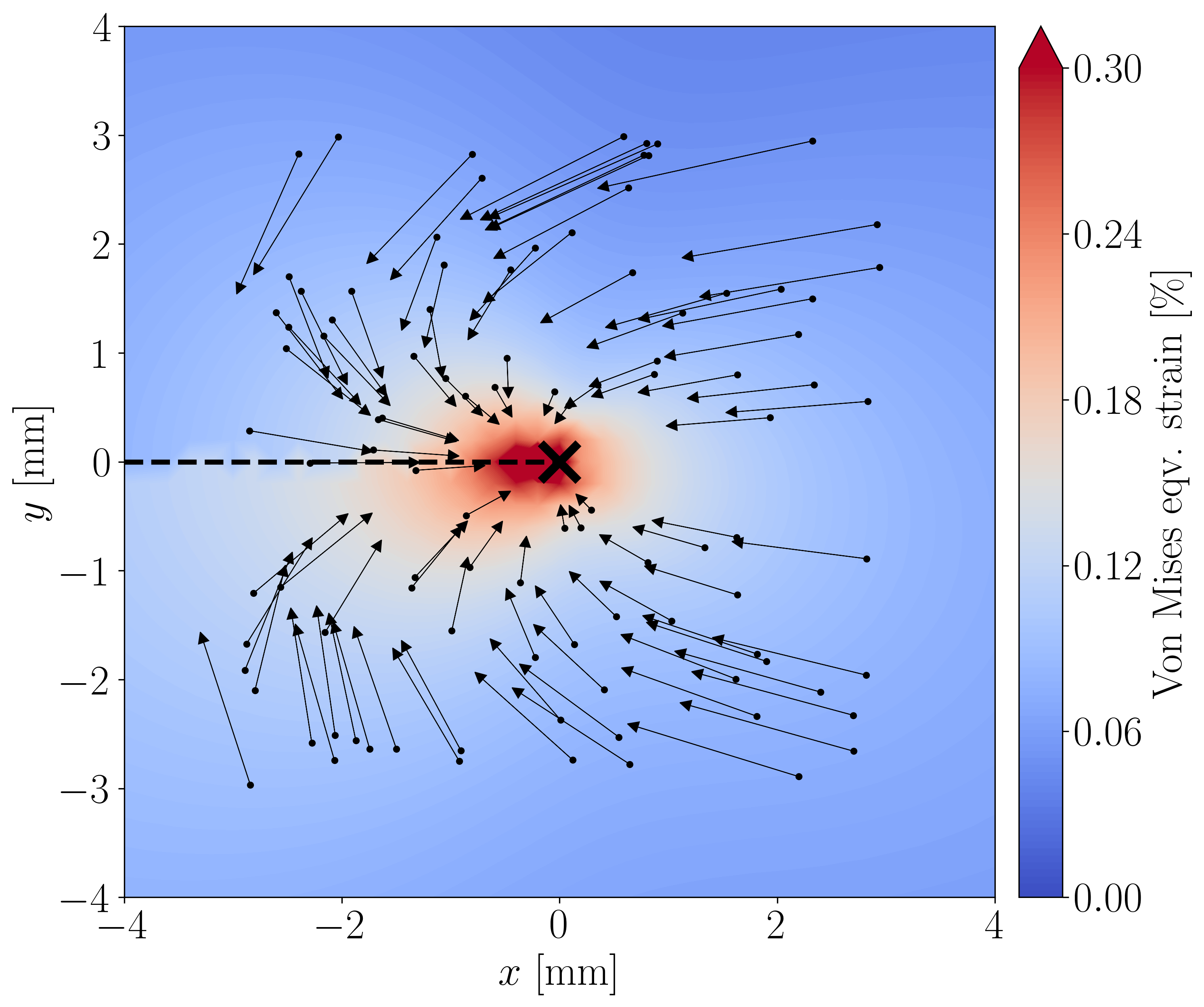}
    \includegraphics[width=0.49\textwidth]{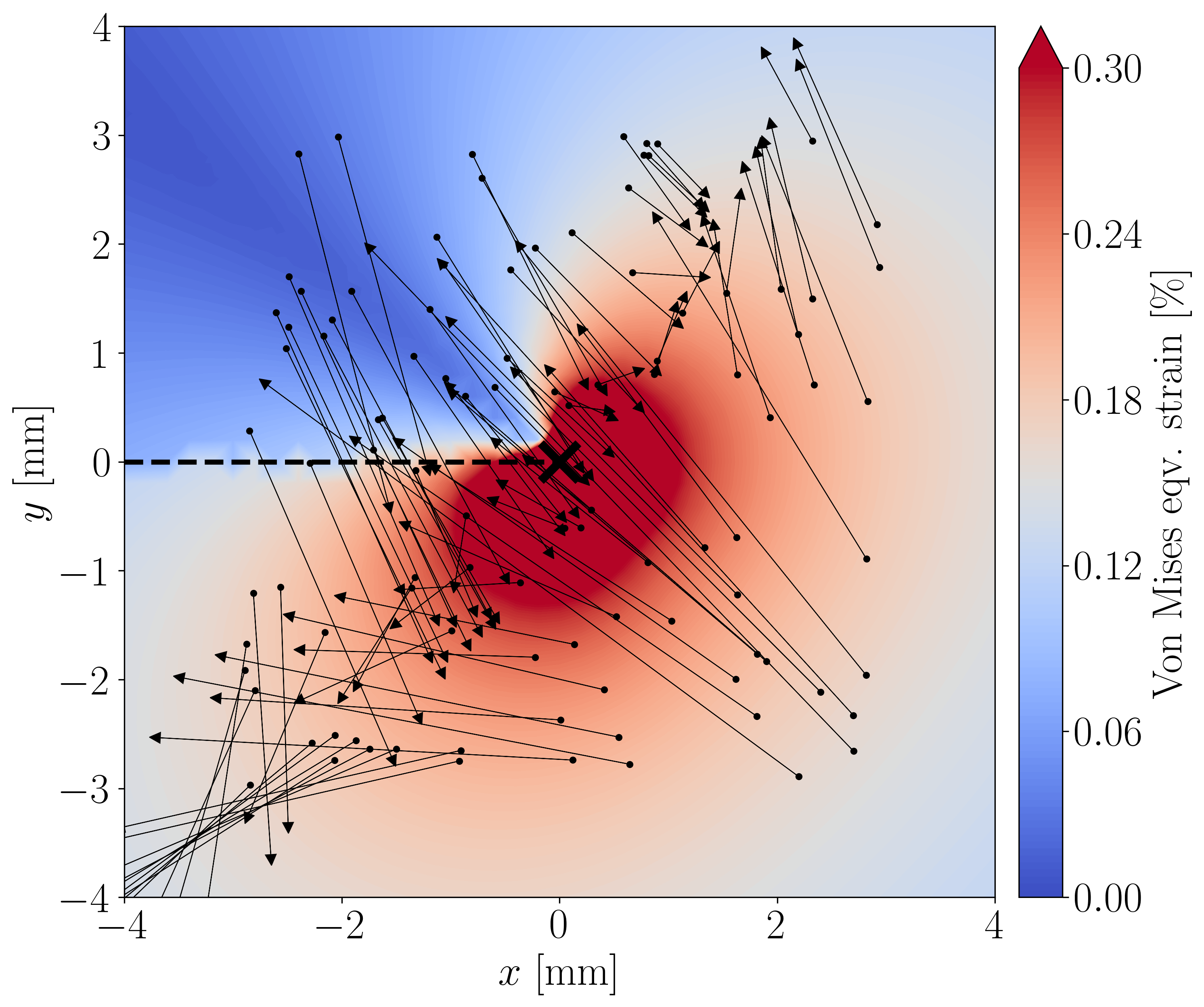}
    \caption{Correction vectors using the mode II formula \eqref{eq:symreg_mode_II} with $c_x^{\rm II} = c_y^{\rm II} = 1$ for the mode II load case $\sigma_{xx} = \sigma_{xy} = 10, \sigma_{yy} = 0$ [MPa] (left) and the mixed-mode load case $\sigma_{xx} = \sigma_{yy} = \sigma_{xy} = 10$ [MPa] (right).}
    \label{fig:vector_plot_mode_II}
\end{figure}

\subsection{Convergence of iterative correction} \label{sec:results_convergence}
In this section, we study the convergence when iterating the discovered crack tip correction formulas until the iteration step size $\sqrt{d_x^2+d_y^2}$ reaches a certain threshold $\delta >0$. Here, we choose $\delta = 10^{-3}$.

In Section \ref{sec:results_vector_fields}, we find that Formula \eqref{eq:symreg_mode_I} is most promising for iterative application in all mode-I-dominated load cases. More complex formulas in Tables \ref{table:eqs_pareto_I} and \ref{table:eqs_pareto_II} have a smaller RMSE meaning that on average they get closer to the crack tip in a single iteration step. However, many of these formulas are unfeasible for iterative application or only work for specific load cases. 
Therefore, we only show the convergence of the iterative correction using Formula \eqref{eq:symreg_mode_I} here. 
However, the iterative behaviour of all pairs of formulas discovered by $\Phi$-SO in Tables \ref{table:eqs_pareto_I}-\ref{table:eqs_pareto_mixed}, is provided as supplementary material to this publication.

\begin{figure}[htbp]
    \centering
    \begin{subfigure}[t]{0.49\textwidth}
        \includegraphics[width=0.95\textwidth]{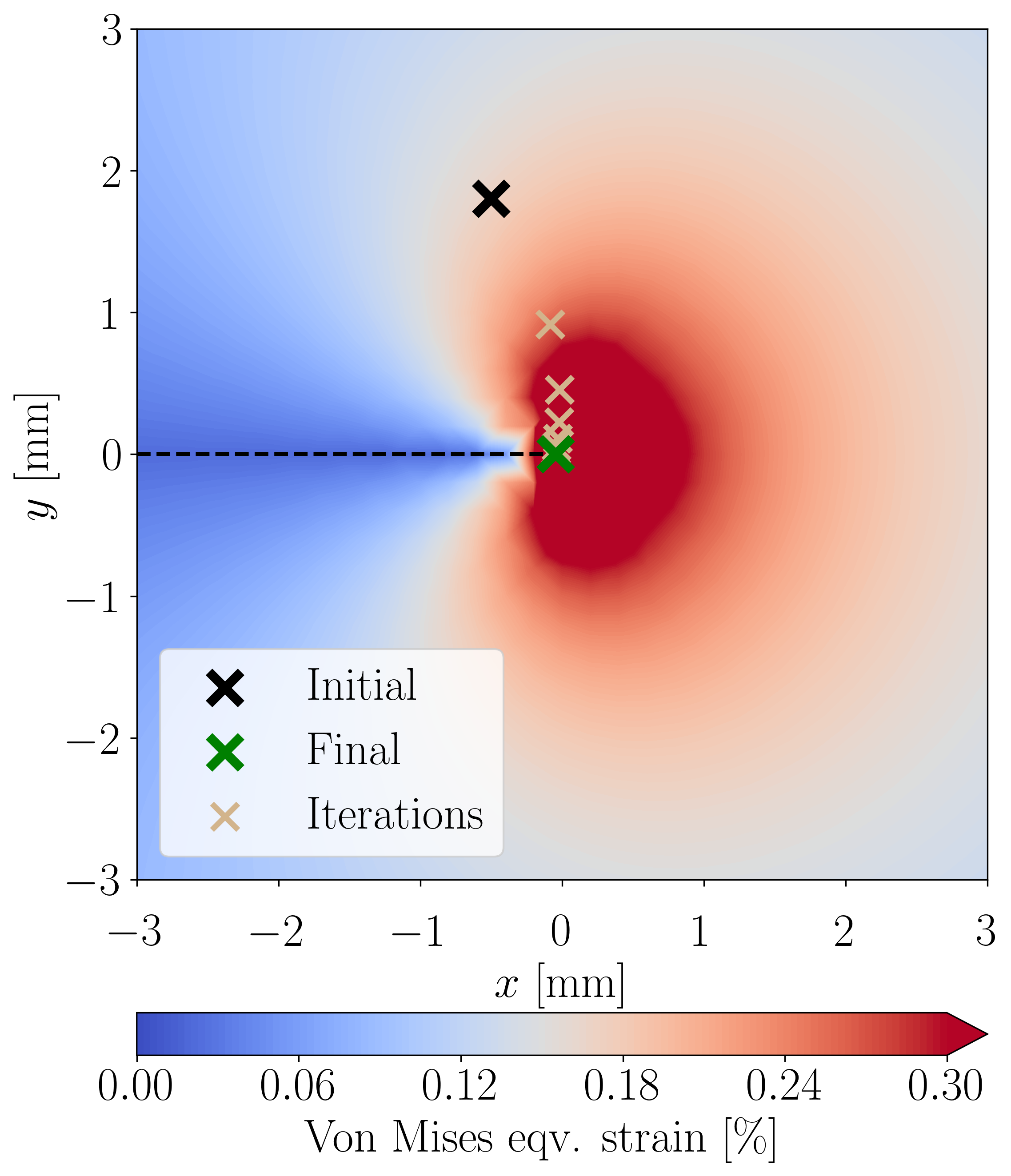}
        \caption{Mode I iterations}
        \label{fig:conv_fem_mode_I_plot}
    \end{subfigure}
    \hfill
    \begin{subfigure}[t]{0.49\textwidth}
        \includegraphics[width=0.95\textwidth]{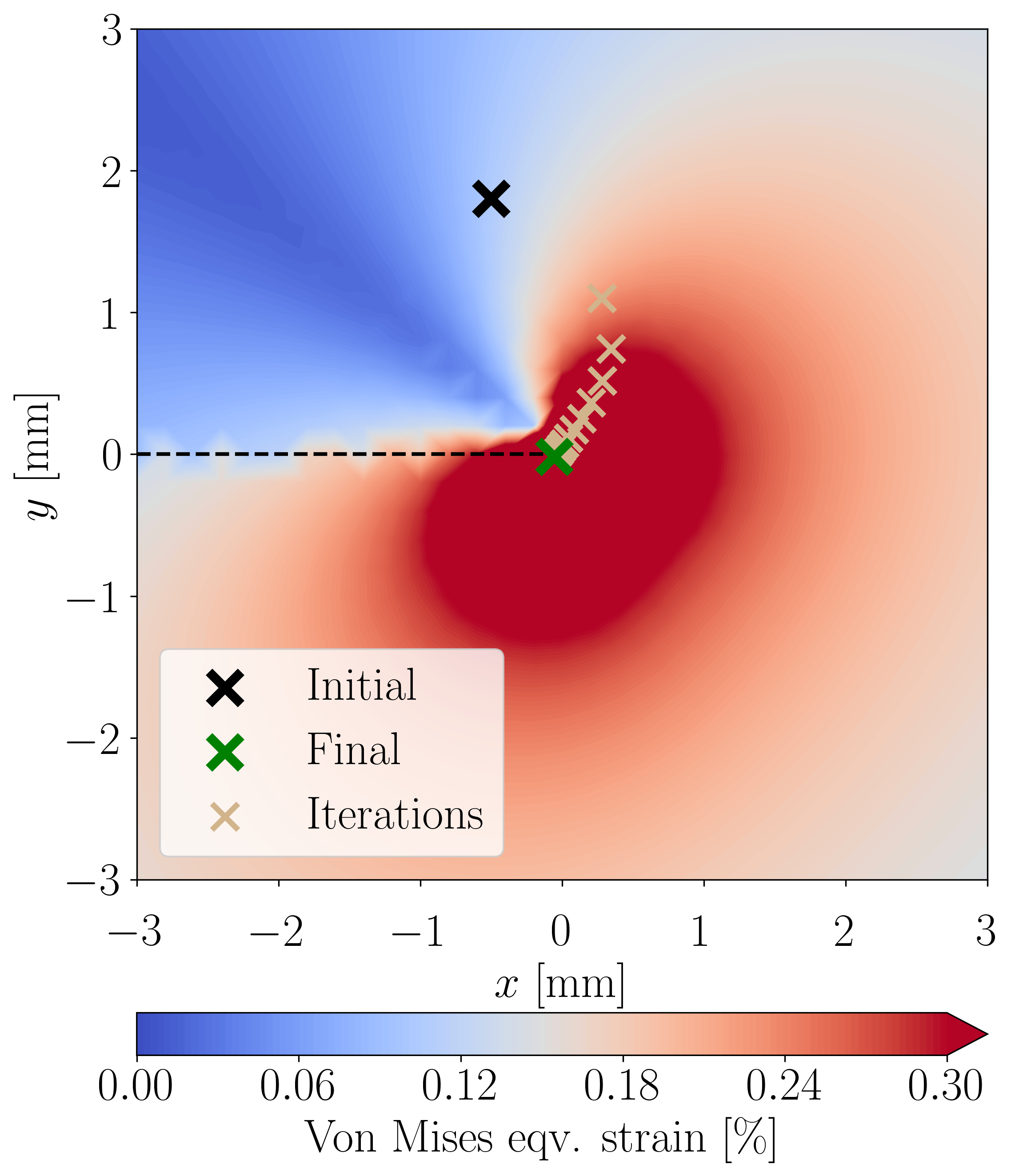}
        \caption{Mixed-mode iterations}
        \label{fig:conv_fem_mixed_mode_plot}
    \end{subfigure}
    \begin{subfigure}[t]{0.49\textwidth}
        \includegraphics[width=0.95\textwidth]{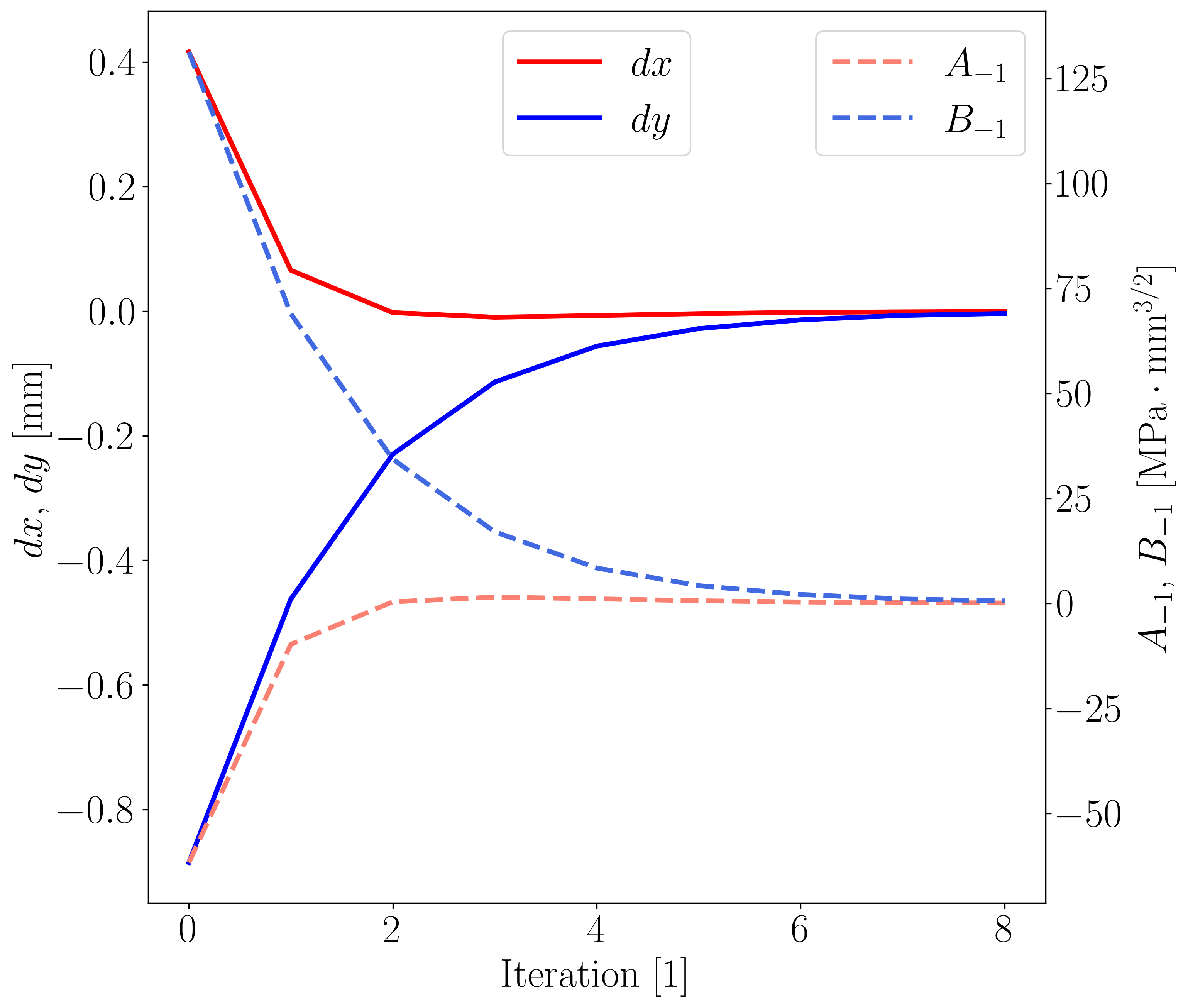}
        \caption{Mode I convergence}
        \label{fig:conv_fem_mode_I_conv}
    \end{subfigure}
    \hfill
    \begin{subfigure}[t]{0.49\textwidth}
        \includegraphics[width=0.95\textwidth]{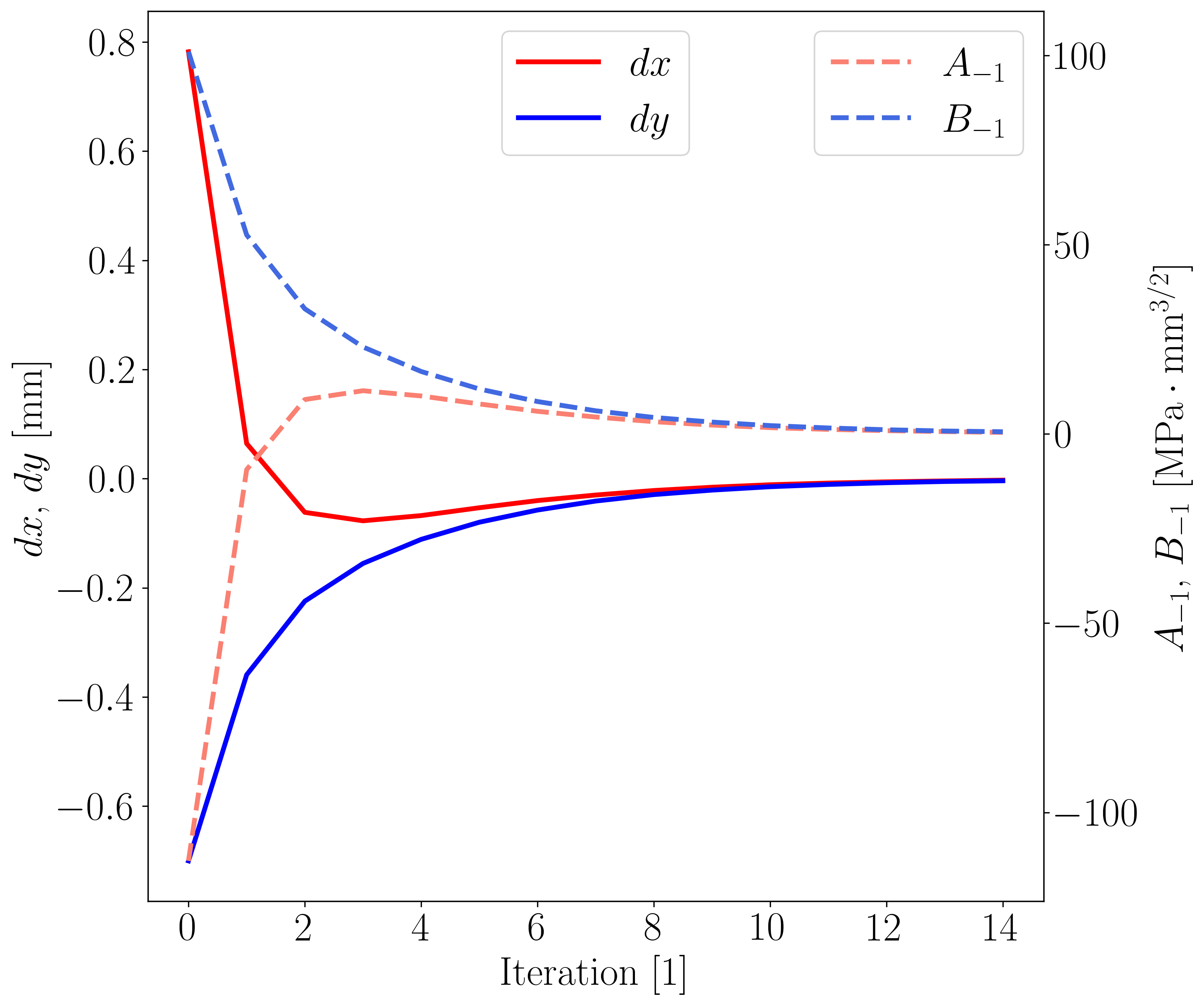}
        \caption{Mixed-mode convergence}
        \label{fig:conv_fem_mixed_mode_conv}
    \end{subfigure}
    \caption{Iterative correction using the mode I formula \eqref{eq:symreg_mode_I} with $c_x^{\rm I} = c_y^{\rm I} = 1$. Top: von Mises eqv. strain with crack tip correction iterations. Bottom: Convergence of $d_x, d_y$ and $A_{-1}, B_{-1}$ to zero. Left: For the mode I load case $\sigma_{xx} = \sigma_{yy} = 10, \sigma_{xy} = 0$ MPa. Right: For the mixed-mode load case $\sigma_{xx} = \sigma_{yy} = \sigma_{xy} = 10$ MPa.}
    \label{fig:conv_fem_mode_I}
\end{figure}

In Figure \ref{fig:conv_fem_mode_I}, we apply iterative correction using the mode I formula \eqref{eq:symreg_mode_I} with $c_x^{\rm I} = c_y^{\rm I} = 1$. Figure \ref{fig:conv_fem_mode_I_plot} and \ref{fig:conv_fem_mixed_mode_plot} show the iteration steps and the final corrections.
As shown in Figure \ref{fig:conv_fem_mode_I_conv} and \ref{fig:conv_fem_mixed_mode_conv}, the algorithm converges under both load cases in a very stable way with $A_{-1}$ and $B_{-1}$ tending to zero accordingly.
The correction takes 8 and 15 iterations to converge to the threshold of $\delta < 10^{\mathrm{-3}}$ for the mode I and mixed-mode example, respectively. In case of FE data, we know the actual crack tip position at $x=y=0$ and can compare it with the final correction. The RMSE of correction is $0.047 \, \mathrm{mm}$ and $0.058 \, \mathrm{mm}$, respectively, which is significantly smaller than the mesh size of the FE model.

\section{Application to experimental data}\label{sec:application}
To show the effectiveness of our method, we apply the discovered crack tip correction formulas to experimental DIC data of growing fatigue cracks captured by two independent optical systems acquiring global full-field and local high-resolution DIC, respectively. The experiments differ in specimen geometry and loading conditions. The tested material in both cases is the aluminium alloy AA2024-T3.

\begin{enumerate}
 \item \textbf{Uniaxial testing of AA2024-T3 sheet material}
 
 Fatigue crack growth experiments of middle tension MT160 specimen of AA2024-T3 were conducted according to ASTM E647-15 \cite{ASTM_E0647}. The basis is a servo-hydraulic testing machine for the sinusoidal fatigue loading with constant amplitude. The maximum load and load ratio were set to $F=15\mathrm{kN}$ and $R=0.1$, respectively. DIC measurements were performed on both sides of the sheet specimen. On one side, a 3D DIC system captures the displacements of the entire surface with a spatial resolution (i.e. facet distance) of $0.59 \, \mathrm{mm}$. On the other side of the specimen, the testing system is equipped with a robot-assisted high-resolution 2D DIC system. It consists of a KUKA LBR iiwa that guides a Zeiss 206C light optical  microscope including a Basler a2A5320-23umPro global shutter 16 Megapixel CMOS camera. It allows the measurement of high resolution DIC displacement fields with a spatial resolution of $0.06 \mathrm{mm}$ of the crack tip region throughout the experiment. A detailed description of the test set up including all algorithms for ensuring good DIC measurement quality is presented by Paysan et al. \cite{Paysan2023}.
 
 \item \textbf{Biaxial testing of AA2024-T3 sheet material} 
 
 According to the experimental setup described in \cite{Breitbarth2018}, a biaxial cyclic load of $F_{\mathrm{max}}=45\,\mathrm{kN}$ and $F_{\mathrm{min}}=4.5\,\mathrm{kN}$, i.e. $R=0.1$, was applied simultaneously in both directions. The crack growth direction, therefore, was perpendicular to the rolling direction of the sheet. The specimen has a thickness of $2.03\,\mathrm{mm}$ and a test field of $420\times420\,\mathrm{mm^2}$. We used the combination of the GOM Aramis 12M DIC system and a robotic arm carrying a light optical  microscope similar to the uniaxial tests. However, both systems monitored the same side of the specimen due to constructional restrictions of the biaxial test rig. The 3D DIC system was focused on a measuring volume of $\approx 500\times 380\,\mathrm{mm^2}$, yielding a spatial resolution of $\approx 8\times8\, \mathrm{pixels/mm^2}$. The light optical  microscope was focussed on a field of view of $28.96\times 16.5\,\mathrm{mm^2}$, allowing a spatial resolution of $\approx 183\times183\, \mathrm{pixels/mm^2}$ using the full camera sensor.
\end{enumerate}

\subsection{Line interception method}\label{sec:lim}
For experimental data obtained by DIC, the challenge of locating an initial starting point for the iterative crack tip correction remains. In addition, the orientation of the crack must be correctly estimated, as the correction formula can only correct the translation, but not the rotation. In this paper, we detect the approximate crack path by fitting a $\tanh$-function to the $y$-displacement on vertical slices perpendicular the the expected crack. We call this \textit{line interception method (LIM)}. 

\begin{figure}[ht]
    \centering
    \includegraphics[width=0.5\textwidth]{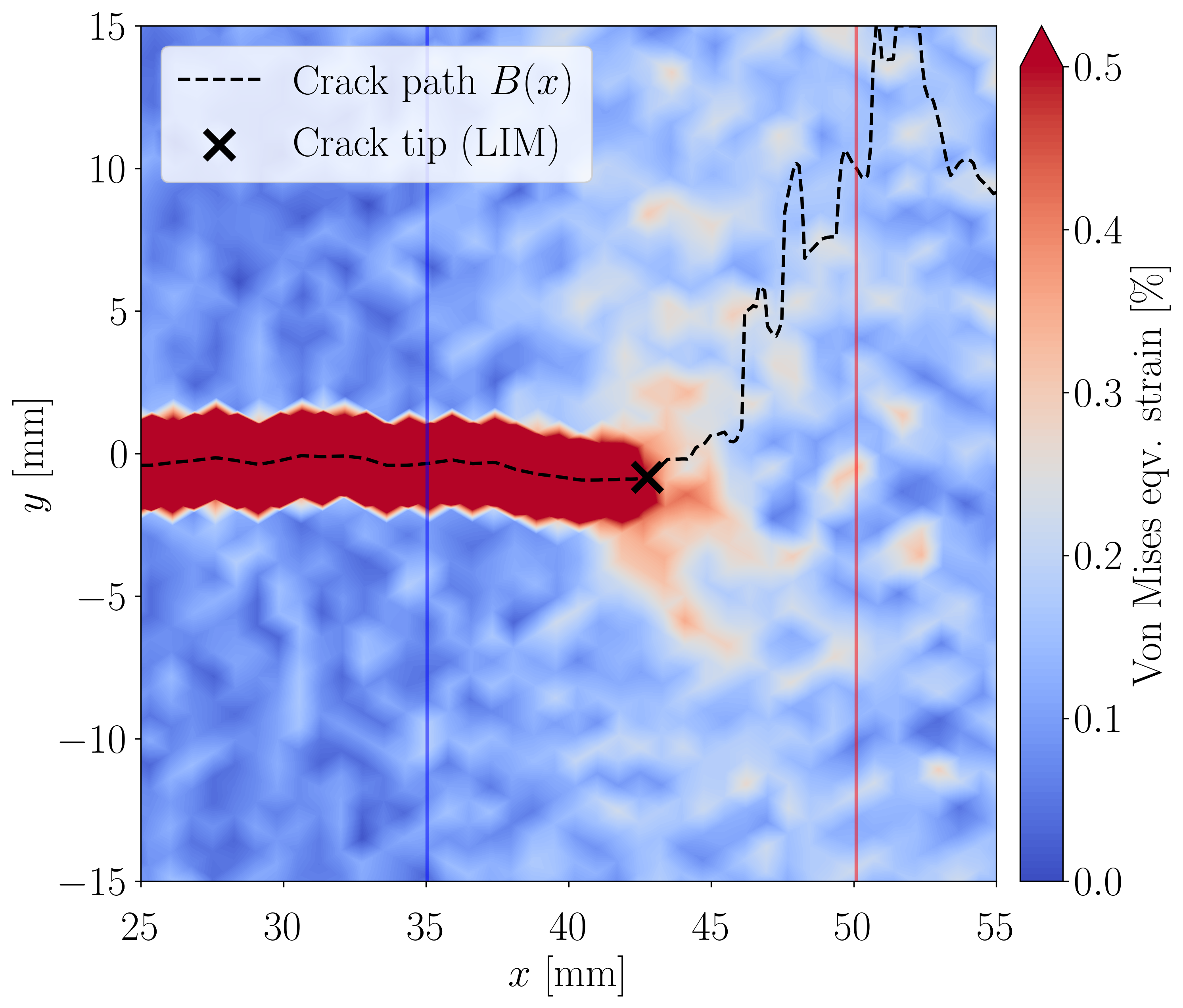}
    \includegraphics[width=0.44\textwidth]{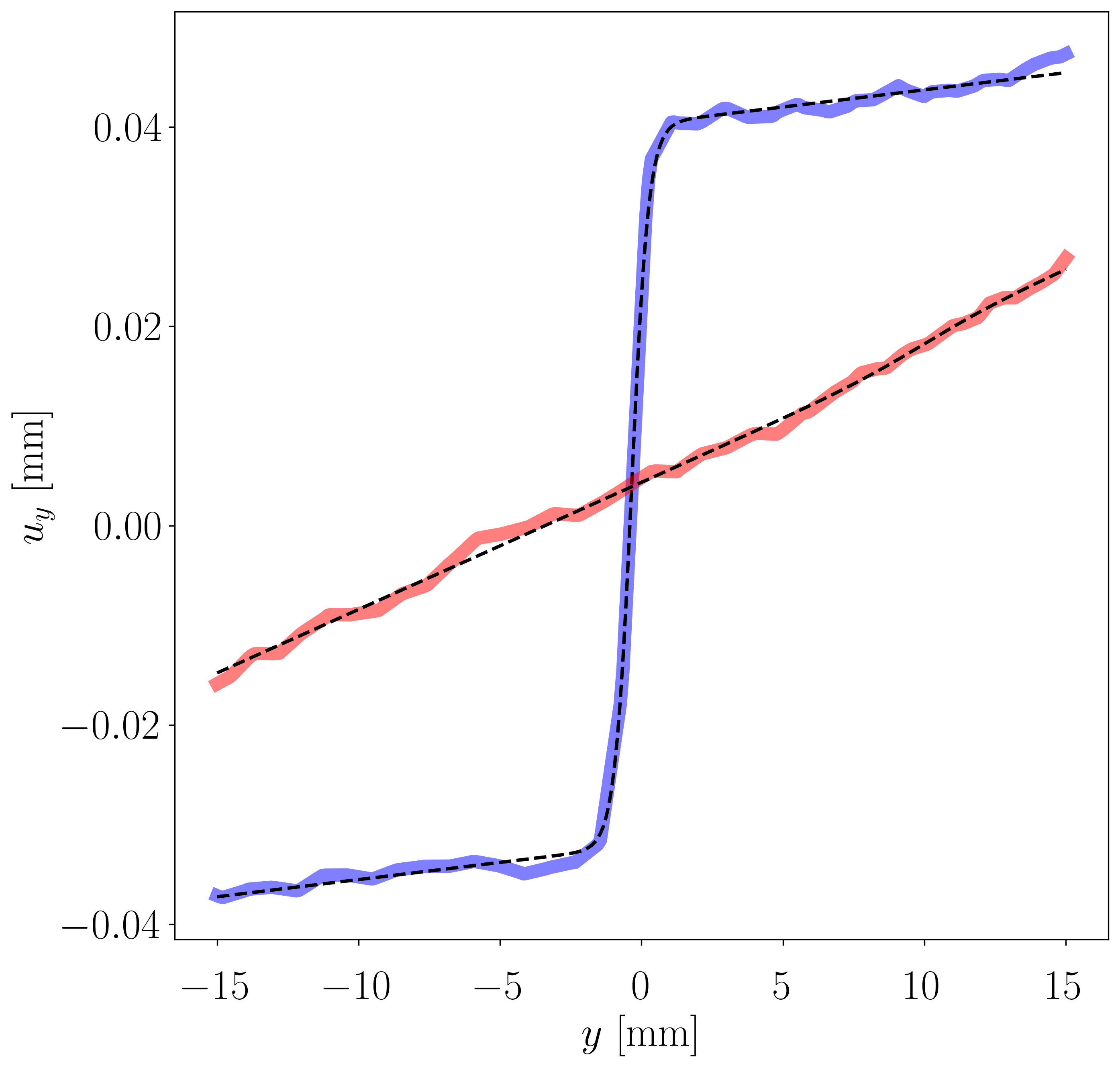}
    \caption{Line interception method (LIM) applied to DIC data. Left: von Mises eqv. strain with estimated crack path and crack tip. Vertical slices $S_x$ at $x\approx 35$ mm (blue) and $x\approx 50$ mm (red). Right: $y$-displacements on these slices overlayed with the fitted $\tanh$ ansatz function (dashed).}
    \label{fig:line_interception_method}
\end{figure}

The implemented LIM builds on displacement gradients perpendicular to the crack path and is mostly insensitive to local scatter. According to Figure \ref{fig:line_interception_method}, we define parallel, equidistant, vertical slices. These slices are roughly perpendicular to the crack path. We interpolate the $y$-displacement onto these slices. When subject to maximum loads, an open crack leads to a steep jump in the $y$-displacement. This characteristic jump is fitted using a $\tanh$ ansatz function, as defined in the following Equation \ref{eq:tanh_fit}. For each slice $S_x$,
\begin{equation} \label{eq:tanh_fit}
    u_y(x, y)=A(x) \cdot \tanh{((y-B(x)) \cdot C(x))} + D(x) \cdot y + E(x)
\end{equation}

For a fixed slice $S_x$, $A(x)$ relates to the distance between the faces of the crack. $B(x)$ is the midpoint between the crack faces, and it is the quantity of interest. $C(x)$ adjusts the slope of the curve. $D(x)$ describes the linear deformation of the base material, whereas $E(x)$ represents a constant offset indicating a rigid-body displacement.

The crack path is reconstructed by plotting $B(x)$ over $x$. However, the gradient $C(x)$ vanishes as the crack tip is approached. Therefore, it is challenging to find the exact crack tip position with this method. We map the equivalent strain onto the reconstructed crack path. By searching for the point, where the strain exceeds a previously defined threshold, we can roughly estimate the crack tip position. Figure \ref{fig:line_interception_method} also shows that the method becomes unstable when a crack is no longer present in the vertical slice $S_x$.

The crack angle is calculated by fitting a line to the LIM-detected crack path ahead of the tip. For this, only the path close to the crack tip should be taken into account. To this end, an \textit{angle estimation radius} needs to be defined. For Figure \ref{fig:line_interception_method} a angle estimation radius of $5 \, \mathrm{mm}$ was used.

We use this approximate crack tip position to initialize the iterative crack tip correction formula and the detected crack angle to correct the crack orientation. It should be noted that any other method to detect the crack path and tip, for instance the machine learning models from \cite{Strohmann2021, Melching2022}, could be used to get an initial guess as well.

\subsection{Iterative correction for high-resolution DIC} \label{sec:application_mDIC}
\begin{figure}[htbp]
    \centering
    \begin{subfigure}[t]{0.49\textwidth}
        \includegraphics[width=0.95\textwidth]{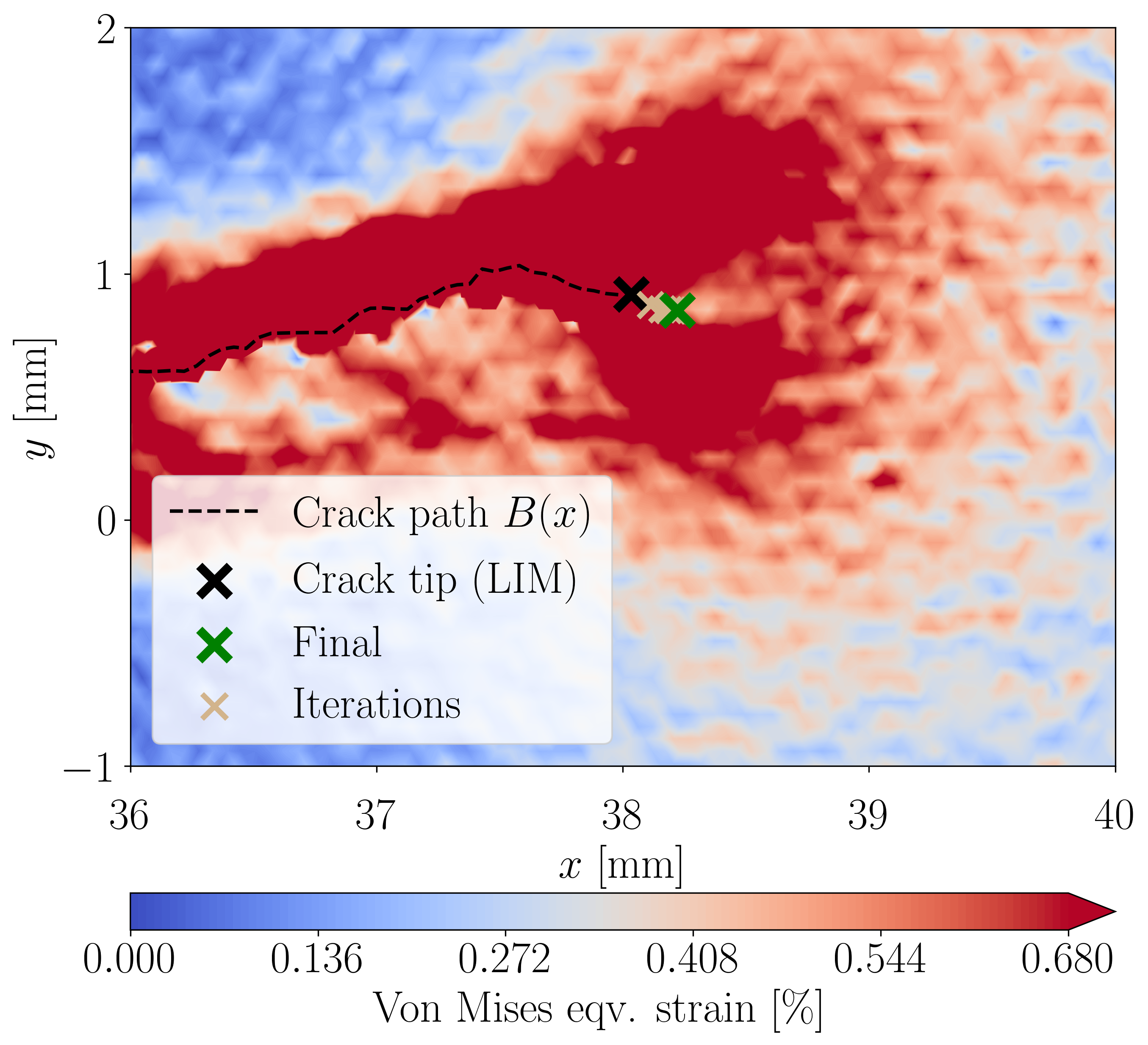}
        \caption{Uniaxial iterations}
        \label{fig:conv_dic_uniaxial_plot}
    \end{subfigure}
    \hfill
    \begin{subfigure}[t]{0.49\textwidth}
        \includegraphics[width=0.95\textwidth]{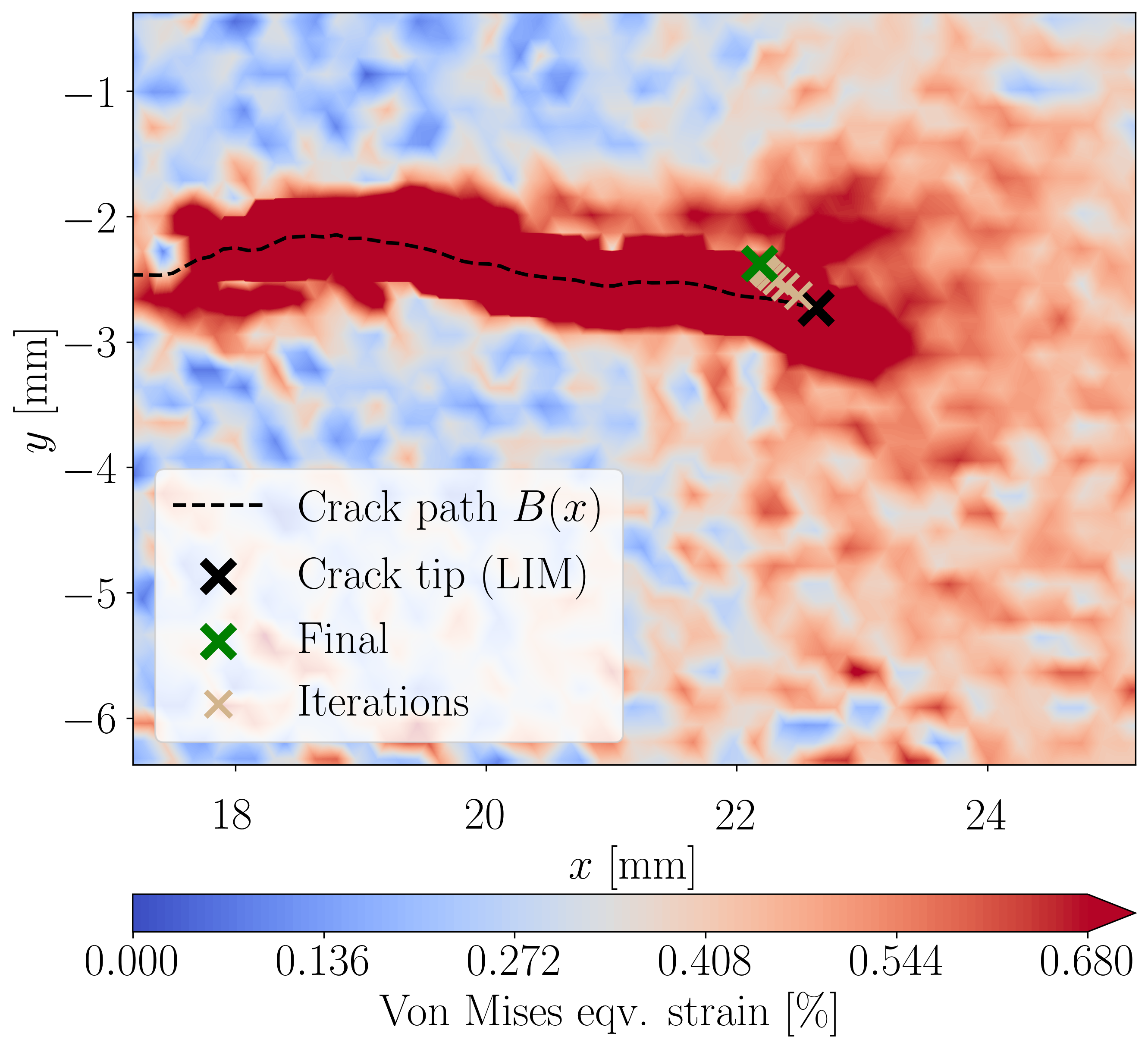}
        \caption{Biaxial iterations}
        \label{fig:conv_dic_biaxial_plot}
    \end{subfigure}
    \begin{subfigure}[t]{0.49\textwidth}
        \includegraphics[width=0.95\textwidth]{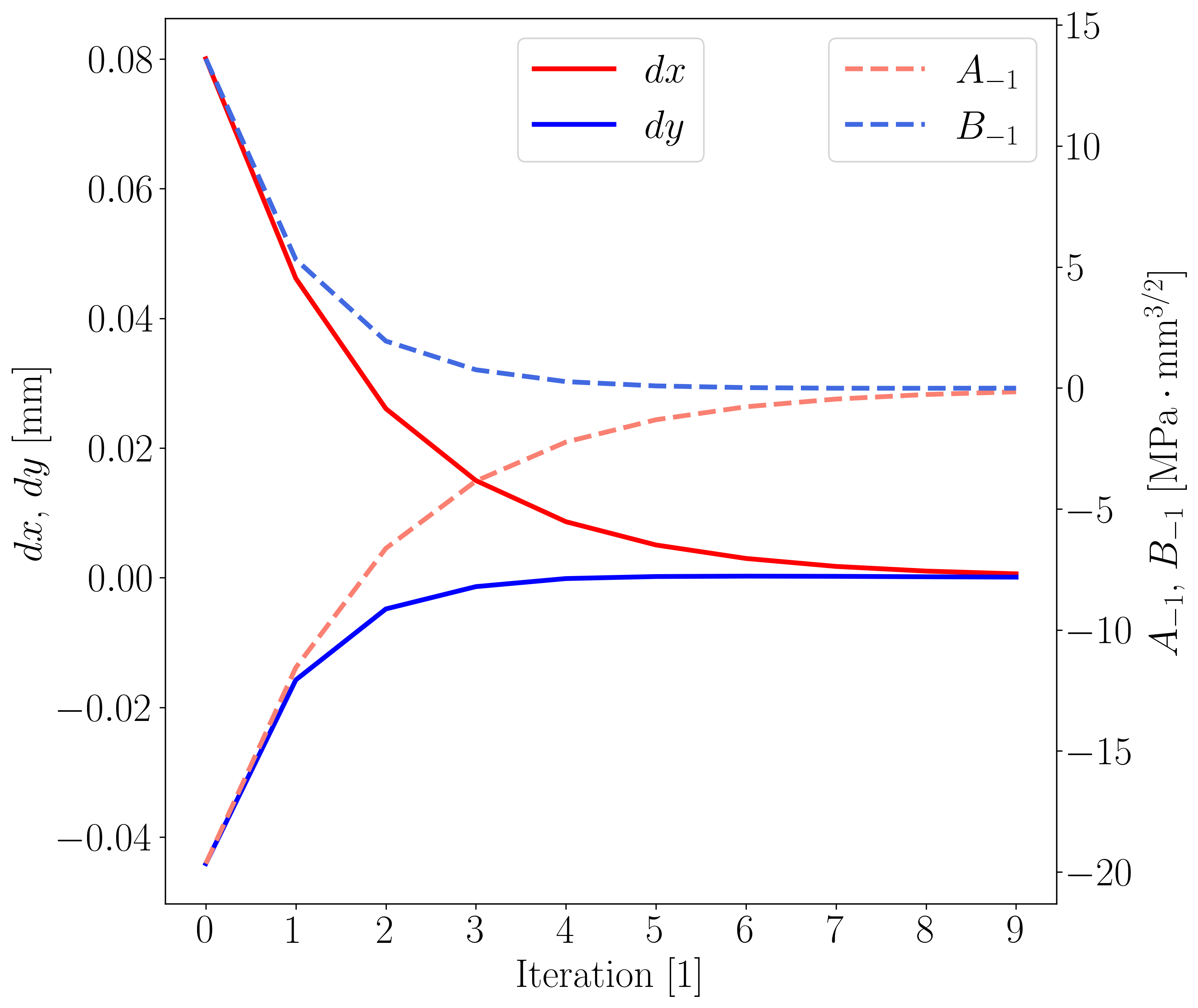}
        \caption{Uniaxial convergence}
        \label{fig:conv_dic_uniaxial_conv}
    \end{subfigure}
    \hfill
    \begin{subfigure}[t]{0.49\textwidth}
        \includegraphics[width=0.95\textwidth]{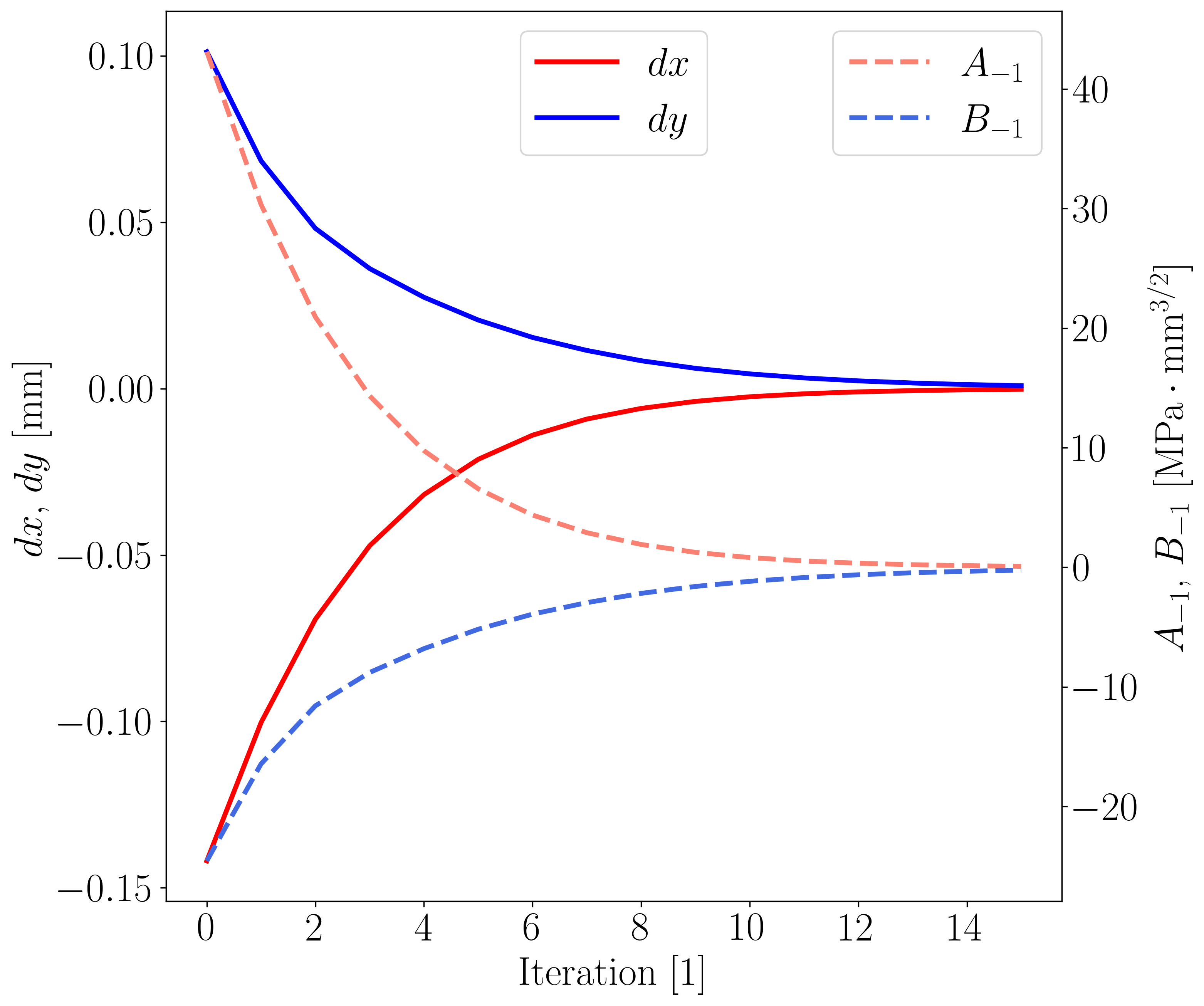}
        \caption{Biaxial convergence}
        \label{fig:conv_dic_biaxial_conv}
    \end{subfigure}
    \caption{Iterative correction for experimental high-resolution DIC data using the mode I formula \eqref{eq:symreg_mode_I} with $c_x^{\rm I} = c_y^{\rm I} = 1$. Top: von Mises eqv. strain with crack tip correction iterations. Bottom: Convergence of $d_x, d_y$ and $A_{-1}, B_{-1}$ to zero. Left: Uniaxial test. Right: Biaxial test.}
    \label{fig:conv_dic}
\end{figure}

The challenging character of experimental data instead of simulated ideal solutions can be seen in Figure \ref{fig:conv_dic}. The figure shows the iterative application of correction Formula \ref{eq:symreg_mode_I} on experimental high-resolution DIC data for one representative time step (crack length) for maximum load during uniaxial (Figure \ref{fig:conv_dic_uniaxial_plot}) and biaxial (Figure \ref{fig:conv_dic_biaxial_plot}) fatigue crack growth experiments, respectively. Along with the inherent scatter, especially high-resolution DIC data provide more challenges for crack tip detection. First, the plastic wake surrounds the crack path with even small branch-like features (see Figure \ref{fig:conv_dic_uniaxial_plot}). Secondly, the DIC system does not necessarily \textit{know} that a crack exists, thus calculating unphysically large strain artifacts very close to the crack path. Third, the crack path can be arbitrarily complex making it difficult to define a single value for the crack angle. 

Although the LIM provides very realistic initial positions for the crack tip, the iterative correction method is preferred because it uses physical knowledge (i.e. the crack tip field) rather than relying solely on raw DIC data. Therefore, we expect $A_{-1}=B_{-1}=0$ when calculating the Williams terms for the true crack tip position as shown by validating the correction method using FE data (see Section \ref{sec:results_convergence}).

For the uniaxial case, we choose $\alpha = 45^\circ$, $r_{\rm min}=1 \, \mathrm{mm}$, $r_{\rm max}=2\, \mathrm{mm}$, and a tick size of $0.02 \, \mathrm{mm}$ for the fitting domain. For the biaxial case, we choose the same tick size and $\alpha$, but move further away from the crack tip ($r_{\rm min}=2\, \mathrm{mm}$, $r_{\rm max}=4\, \mathrm{mm}$).
In Figures \ref{fig:conv_dic_uniaxial_conv} and \ref{fig:conv_dic_biaxial_conv}, the method converges after 10-15 iterations with $A_{-1}$ and $B_{-1}$ tending to zero in both experiments. This confirms the thesis of Baldi and Santucci \cite{Baldi2022}. While the corrected crack tip position for the uniaxial example is consistent with physical intuition, this does not appear to be the case for the biaxial example. Reasons for this could be bifurcation, crack tunneling, or shear lips in $z$-direction.

\subsection{Improved stability}
From a fatigue and lifetime perspective, one is interested in quantifying crack propagation in terms of the crack growth per cycle, $\Delta a/\Delta N$, or incrementally  $\mathrm{d} a/\mathrm{d}N$, with respect to the cyclic stress intensity factor. Measuring the exact crack growth rate by either integral methods such as direct current potential drop or using optical methods can be very challenging since small errors may have a high impact when derivatives are calculated. Typically, results must be averaged to smooth the curves. 

\begin{figure}[htbp]
    \centering
    \begin{subfigure}[t]{0.49\textwidth}
        \includegraphics[width=\textwidth]{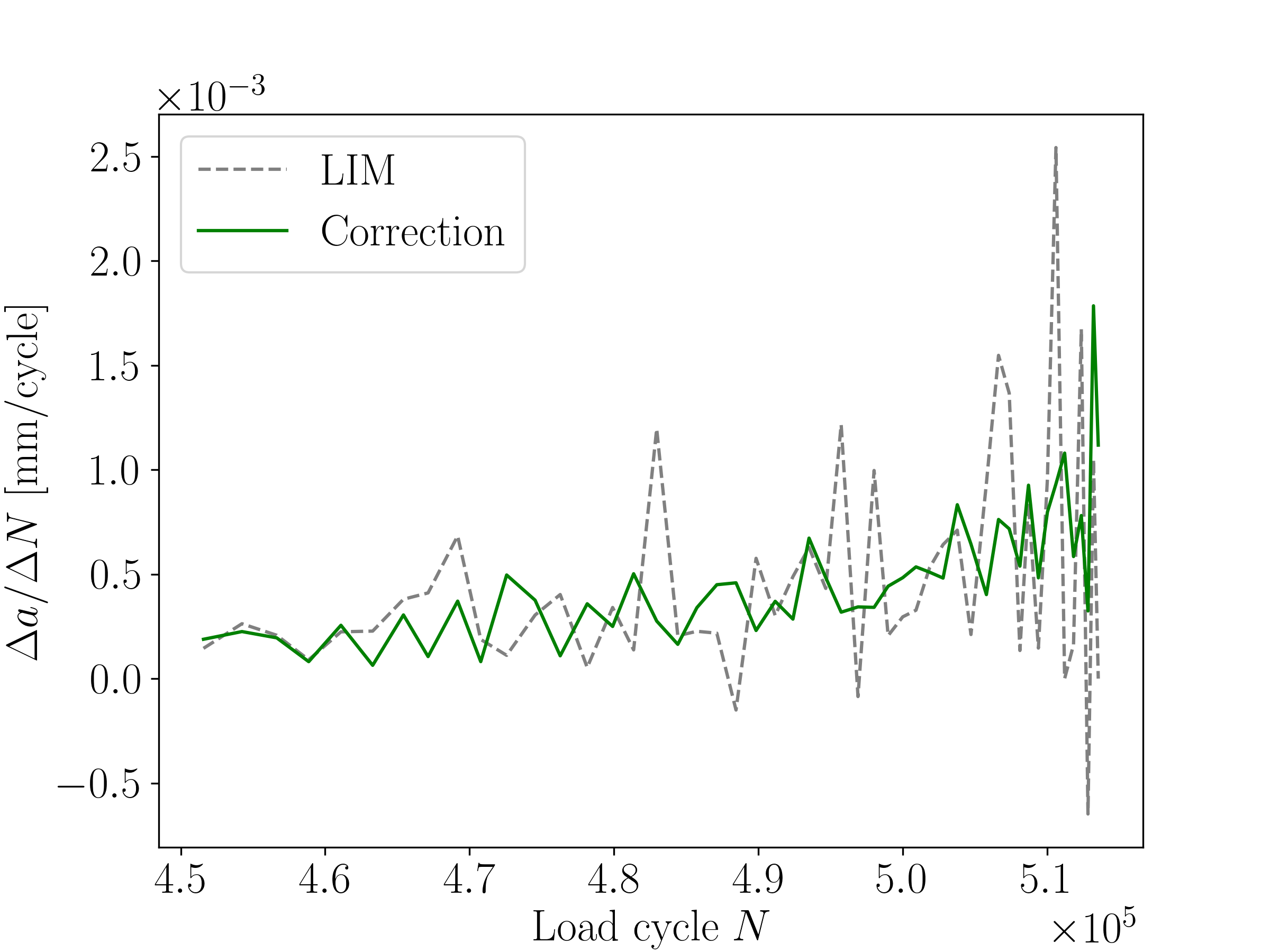}
        \caption{Uniaxial $\Delta a/\Delta N$-stability}
        \label{fig:stability_uniaxial}
    \end{subfigure}
    \hfill
    \begin{subfigure}[t]{0.49\textwidth}
        \includegraphics[width=\textwidth]{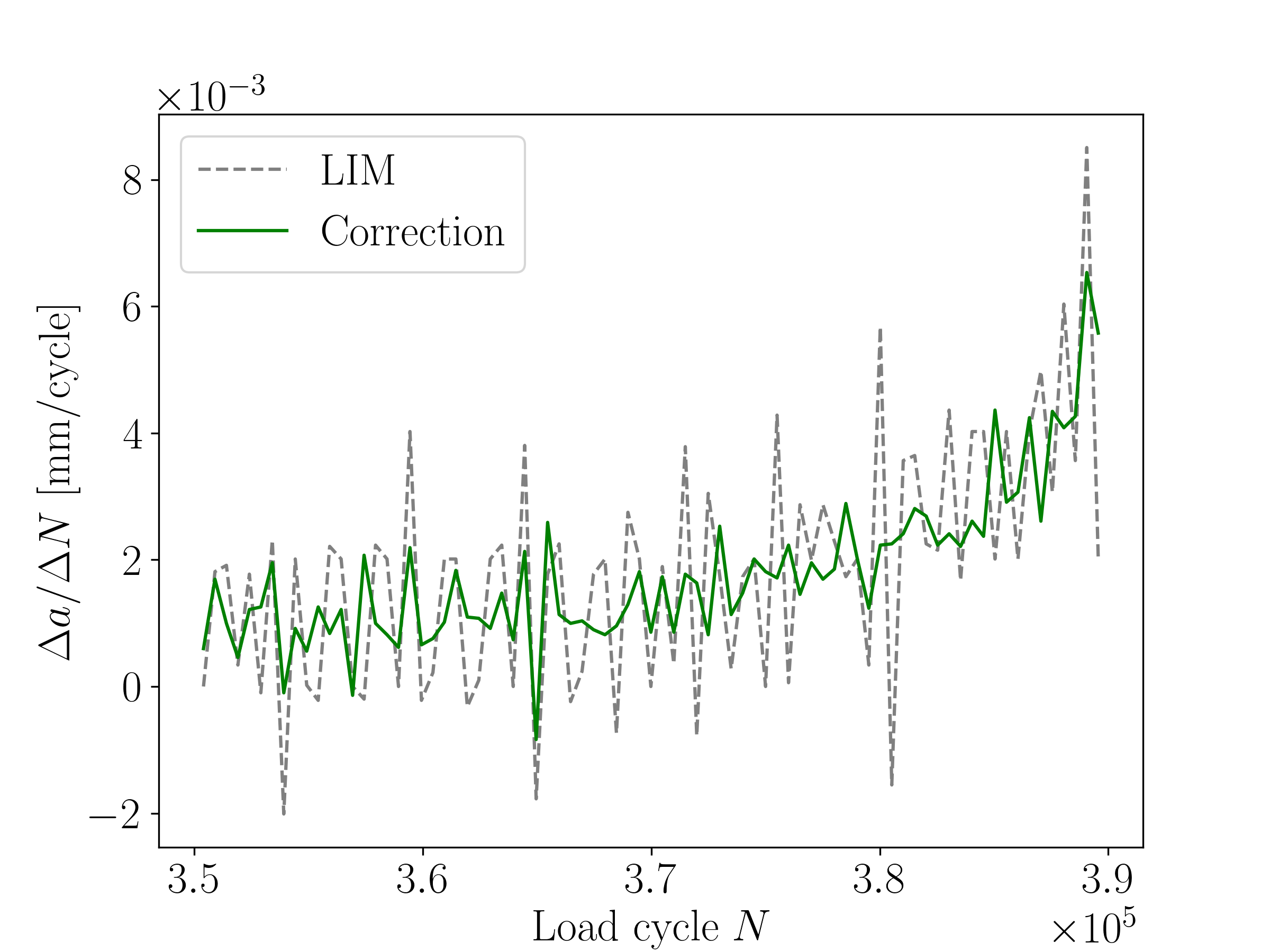}
        \caption{Biaxial $\Delta a/\Delta N$-stability}
        \label{fig:stability_biaxial}
    \end{subfigure}
    \caption{Comparison of crack tip prediction stability between line interception method (LIM) introduced in Section \ref{sec:lim} and correction using our discovered iterative crack tip correction formula \ref{eq:symreg_mode_I}. Left: Uniaxial test of AA2024-T3 sheet material. Right: Biaxial test of AA2024-T3 sheet material (see Section \ref{sec:application})}
    \label{fig:stability}
\end{figure}

In Figure \ref{fig:stability}, we compare the crack growth rates $\Delta a/\Delta N$ calculated as simple differences from successive DIC data using LIM (dashed lines) and the iterative correction method (green line). Results are shown for the uniaxial and biaxial experiment in Figures \ref{fig:stability_uniaxial} and \ref{fig:stability_biaxial}, respectively. We observe that the correction method yields a much smoother curve indicating a lower dependency to errors in the DIC data. While LIM as a threshold method (here, we used a threshold of $\varepsilon_{\mathrm{eqv}}>0.5\,\mathrm{\%}$) is vulnerable to scatter and artifacts in the DIC data, the iterative correction uses the fracture mechanical knowledge, i.e. $A_{\mathrm{-1}}, B_{\mathrm{-1}} \rightarrow 0$, yielding more stable results. This difference between both curves is especially relevant for the number of data points where $\Delta a/\Delta N < 0$ indicating a negative crack growth rate. This is physically impossible. With our iterative correction method, such problematic data points are fully avoided for the uniaxial and reduced to a single data point for the biaxial experiment, respectively. 

\section{Conclusion} \label{sec:conclusion}
We discover an iterative crack tip correction algorithm using deep symbolic regression based on Williams coefficients obtained by crack tip field characterization.

First, we create a diverse data set using simulations of single cracks under mode I, mode II, and mixed-mode loadings. Then, we compute Williams series coefficients at randomly sampled points around the real crack tip position using an over-deterministic fitting approach.
Secondly, we train a deep symbolic regression model with the Williams coefficients $A_{-3},\dots,A_{7}, B_{-3},\dots,B_7$ as input data and the corresponding correction vectors as target output data exploiting physical unit constraints for search space reduction. 
The symbolic regression yields analytical formulas identifying the relevant input variables, i.e. of the 22 input Williams coefficients only $A_{-1},A_{1},B_{-1}$ and $B_{1}$ are used in the most promising formulas. 

We discover distinct correction formulas for mode I, mode II, and mixed-mode load scenarios.
The discovered mode I formula, is a natural extension of \cite{Rethore2015} in the sense that the correction along the crack path direction coincides with the known formula derived by Rethoré. The correction perpendicular to the crack path (in $y$-direction) is new.

While the discovered mode II formula works only under pure mode II loadings, the mode I formula also works under moderate mixed-mode loadings, suggesting a wide applicability for mode-I-dominated experiments.

Finally, we successfully applied this correction formula to experimental digital image correlation data from two different fatigue crack growth experiments - uniaxial and biaxial. 
After a rough estimation of the crack tip using a line interception method, the correction algorithm requires about 5-10 iterations to converge below a correction step size of $10^{-3}$ mm. Simultaneously, the super-singular Williams coefficients $A_{-1}$ and $B_{-1}$ tend to zero.
The correction algorithm improves the stability of the crack tip detection throughout both experiments and works for global full-field DIC as well as local high-resolution DIC.

\section{Acknowledgements}
We acknowledge the financial support of the DLR-Directorate Aeronautics. This work was supported by the Deutsche Forschungsgemeinschaft, Germany (DFG) via the project Experimental analysis and phase-field modelling of the interaction between plastic zone and fatigue crack growth in ductile materials under complex loading (grant number BR 6259/2-1). Furthermore, funding came from the Federal Ministry for Economic Affairs and Climate Action, Germany on the basis of a decision by the German Bundestag, within the aerospace programme LuFo-VI of the project "Untersuchung der Prozess-Struktur-Eigenschaftsbeziehungen rollgeformter und rollprofilierter Profile für eine kosten- und ökooptimierte Türumgebungsstruktur" (Funding ID 20W2102C).

\section{Data availability}
The code and data will be made publicly available on Github (\url{https://github.com/dlr-wf}) and Zenodo (\url{https://doi.org/10.5281/zenodo.10730749}).

\section{Competing interests}
The Authors declare no competing financial or non-financial interests.

\section{Author contributions}
D.M. and E.B. designed the methodology of learning crack tip correction formulas with symbolic regression, generated the training data set, trained and evaluated the symbolic regression models.
F.P. and T.S. designed the high-fidelity testing framework and conducted the experiments. All authors discussed, analysed, and interpreted the results and wrote the manuscript.

\printbibliography

\newpage
\section*{Appendix}

\subsection*{All discovered formulas}
For clarity and brevity, we skipped repetitions and only highlighted the most relevant Pareto formulas in Tables \ref{table:eqs_pareto_I}-\ref{table:eqs_pareto_mixed}. For sake of completeness, we report all formulas discovered by the $\Phi$-SO tool in Tables \ref{table:appendix_eqs_pareto_I}-\ref{table:appendix_eqs_pareto_mixed} below.

\begin{table}[htbp!]
\centering
 \begin{tabular}{|c c c c c|} 
\hline
\multicolumn{5}{|c|}{Parallel correction - $d_x$} \\
\hline
\# & Complexity & Reward & RMSE & Equation \\ [0.5ex]
\hline
0 & 4 &  0.66272 &  0.89169 & $- \frac{A_{-1}}{A_{1}}$\\
1 & 5 &  0.82885 &  0.36178 & $- \frac{1.90537 A_{-1}}{A_{1}}$\\
2 & 7 &  0.86337 &  0.27726 & $- 0.23575 \mathrm{mm} - \frac{1.94928 A_{-1}}{A_{1}}$\\
3 & 8 &  0.86337 &  0.27726 & $- 0.23575 \mathrm{mm} - \frac{1.94928 A_{-1}}{A_{1}}$\\
4 & 9 &  0.86337 &  0.27726 & $- 0.23575 \mathrm{mm} - \frac{1.94928 A_{-1}}{A_{1}}$\\
5 & 12 &  0.86364 &  0.27665 & $- 0.22136 \mathrm{mm} - \frac{1.95236}{\frac{A_{1}}{A_{-1}} + \frac{A_{3}}{A_{1}}}$\\
6 & 14 &  0.86367 &  0.27656 & $- 0.20757 \mathrm{mm} - \frac{A_{1}}{\frac{0.51161 A_{1}^{2}}{A_{-1}} + A_{3}}$\\
7 & 15 &  0.87618 &  0.24759 & $- 0.15055 \mathrm{mm} - \frac{7.64174}{\frac{2.0 A_{1}}{A_{-1}} - \frac{A_{-1}}{A_{-3}}}$\\
8 & 16 &  0.88692 &  0.22339 & $- 0.18292 \mathrm{mm} - \frac{11.66037}{\frac{4.0 A_{1}}{A_{-1}} - \frac{A_{-1}}{A_{-3}}}$\\
9 & 17 &  0.88692 &  0.22339 & $- 0.18293 \mathrm{mm} - \frac{11.66051}{\frac{4.0 A_{1}}{A_{-1}} - \frac{A_{-1}}{A_{-3}}}$\\
10 & 18 &  0.88692 &  0.22339 & $- 0.18292 \mathrm{mm} - \frac{11.66052}{\frac{4.0 A_{1}}{A_{-1}} - \frac{A_{-1}}{A_{-3}}}$\\
11 & 19 &  0.88703 &  0.22314 & $- 0.18589 \mathrm{mm} - \frac{2.84792}{\frac{A_{1}}{A_{-1}} - \frac{0.23186 A_{-1}}{A_{-3}}}$\\
12 & 20 &  0.91267 &  0.16766 & $- 0.00193 \mathrm{mm} - \frac{0.00276 A_{-2} \mathrm{mm}}{\mathrm{N}} - \frac{2.08617 A_{-1}}{A_{1}}$\\
\hline
\hline
\multicolumn{5}{|c|}{Perpendicular correction - $d_y$} \\
\hline
\# & Complexity & Reward & RMSE & Equation \\ [0.5ex]
\hline
0 & 4 &  0.61570 &  1.09381 & $- \frac{B_{-1}}{A_{1}}$\\
1 & 5 &  0.82705 &  0.36646 & $- \frac{2.50796 B_{-1}}{A_{1}}$\\
2 & 7 &  0.82775 &  0.36465 & $0.03639 \mathrm{mm} - \frac{2.5104 B_{-1}}{A_{1}}$\\
3 & 8 &  0.82775 &  0.36465 & $\frac{0.03639 A_{1} \mathrm{mm} - B_{-1} \cdot \left(2.5104 - i \pi\right)}{A_{1}}$\\
4 & 11 &  0.82882 &  0.36193 & $- \frac{73.82141 B_{-1} \mathrm{mm}}{29.36224 \left|{A_{1}}\right| \mathrm{mm} + B_{-1}}$\\
5 & 12 &  0.83067 &  0.35723 & $- 2.99333 \mathrm{mm} + \left|{2.99333 \mathrm{mm} - \frac{2.56666 B_{-1}}{A_{1}}}\right|$\\
6 & 13 &  0.96146 &  0.07025 & $\frac{9.78182 B_{1} \mathrm{mm} - 1.66341 B_{-1}}{A_{1}}$\\
\hline
\end{tabular}
\caption{Pareto formulas - Mode I correction}
\label{table:appendix_eqs_pareto_I}
\end{table}

\begin{table}[htbp!]
\centering
 \begin{tabular}{|c c c c c|} 
\hline
\multicolumn{5}{|c|}{Parallel correction - $d_x$} \\
\hline
\# & Complexity & Reward & RMSE & Equation \\ [0.5ex]
\hline
0 & 4 &  0.57814 &  1.27848 & $- \frac{B_{-1}}{B_{1}}$\\
1 & 5 &  0.60438 &  1.14692 & $1.92379 \mathrm{mm} - \frac{B_{0}}{B_{2}}$\\
2 & 6 &  0.65980 &  0.90339 & $2.30843 \mathrm{mm} - \left|{\frac{B_{0}}{B_{2}}}\right|$\\
3 & 7 &  0.71068 &  0.71328 & $1.2532 \mathrm{mm} - \frac{0.6405 B_{0}}{B_{2}}$\\
4 & 8 &  0.78389 &  0.48304 & $17.06448 \mathrm{mm} - \frac{0.12296 B_{1}^{2}}{B_{2}^{2}}$\\
5 & 9 &  0.78389 &  0.48304 & $17.06447 \mathrm{mm} - \frac{0.12296 B_{1}^{2}}{B_{2}^{2}}$\\
6 & 12 &  0.81061 &  0.40936 & $13.27914 \mathrm{mm} - \frac{0.09304 \left(B_{0} B_{2} + B_{1}^{2}\right)^{2}}{B_{1}^{2} B_{2}^{2}}$\\
7 & 14 &  0.81061 &  0.40936 & $13.27914 \mathrm{mm} - \frac{0.09304 \left(B_{0} B_{2} + B_{1}^{2}\right)^{2}}{B_{1}^{2} B_{2}^{2}}$\\
8 & 15 &  0.81112 &  0.40799 & $13.83444 \mathrm{mm} - \frac{0.09642 \left(B_{1} - B_{2} \left|{\frac{B_{0}}{B_{1}}}\right|\right)^{2}}{B_{2}^{2}}$\\
9 & 17 &  0.81229 &  0.40488 & $12.18511 \mathrm{mm} - \frac{0.08299 B_{1}^{2}}{B_{2}^{2}} - \frac{0.28808}{\left|{\frac{B_{2}}{B_{0}}}\right|}$\\
10 & 18 &  0.84853 &  0.31276 & $- \frac{2.43992 B_{-1}}{B_{1}} + \frac{0.27118 B_{-2}}{B_{2} \mathrm{mm}}$\\
\hline
\hline
\multicolumn{5}{|c|}{Perpendicular correction - $d_y$} \\
\hline
\# & Complexity & Reward & RMSE & Equation \\ [0.5ex]
\hline
0 & 4 &  0.59518 &  1.19191 & $- \frac{A_{-1}}{B_{1}}$\\
1 & 5 &  0.73554 &  0.63006 & $- \frac{2.62186 A_{-1}}{B_{1}}$\\
2 & 6 &  0.86479 &  0.27400 & $- \frac{2.29023 A_{-3}}{B_{1} \mathrm{mm}}$\\
3 & 7 &  0.86479 &  0.27400 & $- \frac{2.29022 A_{-3}}{B_{1} \mathrm{mm}}$\\
4 & 8 &  0.86479 &  0.27400 & $- \frac{2.29025 A_{-3}}{B_{1} \mathrm{mm}}$\\
5 & 9 &  0.87771 &  0.24416 & $\frac{1.0 A_{-1} \mathrm{mm} - 3.08621 A_{-3}}{B_{1} \mathrm{mm}}$\\
6 & 11 &  0.87885 &  0.24158 & $\frac{0.78473 A_{-1} \mathrm{mm} - 2.91486 A_{-3}}{B_{1} \mathrm{mm}}$\\
7 & 12 &  0.88124 &  0.23615 & $- 0.06204 \mathrm{mm} + \frac{A_{-1}}{B_{1}} - \frac{3.08688 A_{-3}}{B_{1} \mathrm{mm}}$\\
8 & 16 &  0.88124 &  0.23615 & $- 0.06204 \mathrm{mm} + \frac{A_{-1}}{B_{1}} - \frac{3.08688 A_{-3}}{B_{1} \mathrm{mm}}$\\
9 & 19 &  0.88148 &  0.23561 & $- \frac{2.19136 A_{7} A_{-3} \mathrm{mm}^{2}}{B_{1}^{2}}$\\
\hline
\end{tabular}
\caption{Pareto formulas - Mode II correction}
\label{table:appendix_eqs_pareto_II}
\end{table}

\begin{table}[htbp!]
\centering
 \begin{tabular}{|c c c c c|} 
\hline
\multicolumn{5}{|c|}{Parallel correction - $d_x$} \\
\hline
\# & Complexity & Reward & RMSE & Equation \\ [0.5ex]
\hline
0 & 4 &  0.51651 &  1.64001 & $\frac{1.0 \cdot 10^{-5} B_{-3}^{2}}{\mathrm{N}^{2}}$\\
1 & 5 &  0.54302 &  1.47437 & $\frac{0.05626 A_{4}}{A_{6}}$\\
2 & 6 &  0.56735 &  1.33602 & $- \frac{A_{-1}}{A_{1} + B_{1}}$\\
3 & 8 &  0.61493 &  1.09711 & $\frac{1.0 \cdot 10^{-5} A_{-1} \mathrm{mm}}{\left|{A_{-1}}\right|}$\\
4 & 9 &  0.61716 &  1.08681 & $- \frac{0.00735 A_{-1}}{A_{1}}$\\
5 & 12 &  0.61971 &  1.07510 & $- 0.18524 \mathrm{mm} + \frac{0.00749 A_{-1}}{A_{1}}$\\
6 & 13 &  0.61971 &  1.07510 & $- 0.18524 \mathrm{mm} + \frac{0.00749 A_{-1}}{A_{1}}$\\
7 & 14 &  0.62085 &  1.06992 & $- 0.16233 \mathrm{mm} - \frac{0.00033 A_{-1}}{A_{1}}$\\
8 & 15 &  0.62179 &  1.06566 & $- 0.18905 \mathrm{mm} + \frac{0.00755 A_{-1}}{A_{1}}$\\
9 & 16 &  0.62215 &  1.06402 & $3.14533 \mathrm{mm} - \left|{3.14533 \mathrm{mm} + \frac{0.00814 A_{-1}}{A_{1}}}\right|$\\
10 & 17 &  0.62297 &  1.06032 & $\frac{0.0153 A_{-1}}{A_{1}}$\\
11 & 18 &  0.64105 &  0.98101 & $\frac{0.0429 A_{1} A_{-1}}{A_{3} \cdot \left(2.88427 A_{1} \mathrm{mm} - A_{-1}\right)}$\\
\hline
\hline
\multicolumn{5}{|c|}{Perpendicular correction - $d_y$} \\
\hline
\# & Complexity & Reward & RMSE & Equation \\ [0.5ex]
\hline
0 & 4 &  0.50001 &  1.75226 & $- 0.0421 \mathrm{mm}$\\
1 & 5 &  0.50660 &  1.70664 & $0$\\
2 & 6 &  0.56391 &  1.35510 & $- \frac{B_{-1}}{A_{1} + B_{1}}$\\
3 & 8 &  0.59950 &  1.17065 & $- \frac{1.0 \cdot 10^{-5} B_{-1} \mathrm{mm}}{\left|{B_{-1}}\right|}$\\
4 & 12 &  0.59954 &  1.17044 & $- \frac{0.0001 B_{-1}}{A_{1}}$\\
5 & 13 &  0.60676 &  1.13566 & $\frac{8.0 \cdot 10^{-5} B_{-1}}{A_{1}}$\\
6 & 14 &  0.60688 &  1.13507 & $\frac{9.0 \cdot 10^{-5} B_{-1}}{\left|{A_{1}}\right|}$\\
7 & 16 &  0.60701 &  1.13449 & $- \frac{9.0 \cdot 10^{-5} B_{-1}}{A_{1}}$\\
8 & 18 &  0.62042 &  1.07207 & $\frac{0.02276 B_{-1}}{A_{3} \mathrm{mm}}$\\
9 & 19 &  0.62106 &  1.06919 & $- 1.0 \cdot 10^{-5} \mathrm{mm} - \frac{0.02166 B_{-1}}{\left|{A_{3}}\right| \mathrm{mm}}$\\
\hline
\end{tabular}
\caption{Pareto formulas - Mixed-mode correction}
\label{table:appendix_eqs_pareto_mixed}
\end{table}

\newpage
\subsection*{Additional vector plots}
\begin{figure}[htbp]
    \centering
    \includegraphics[width=0.49\textwidth]{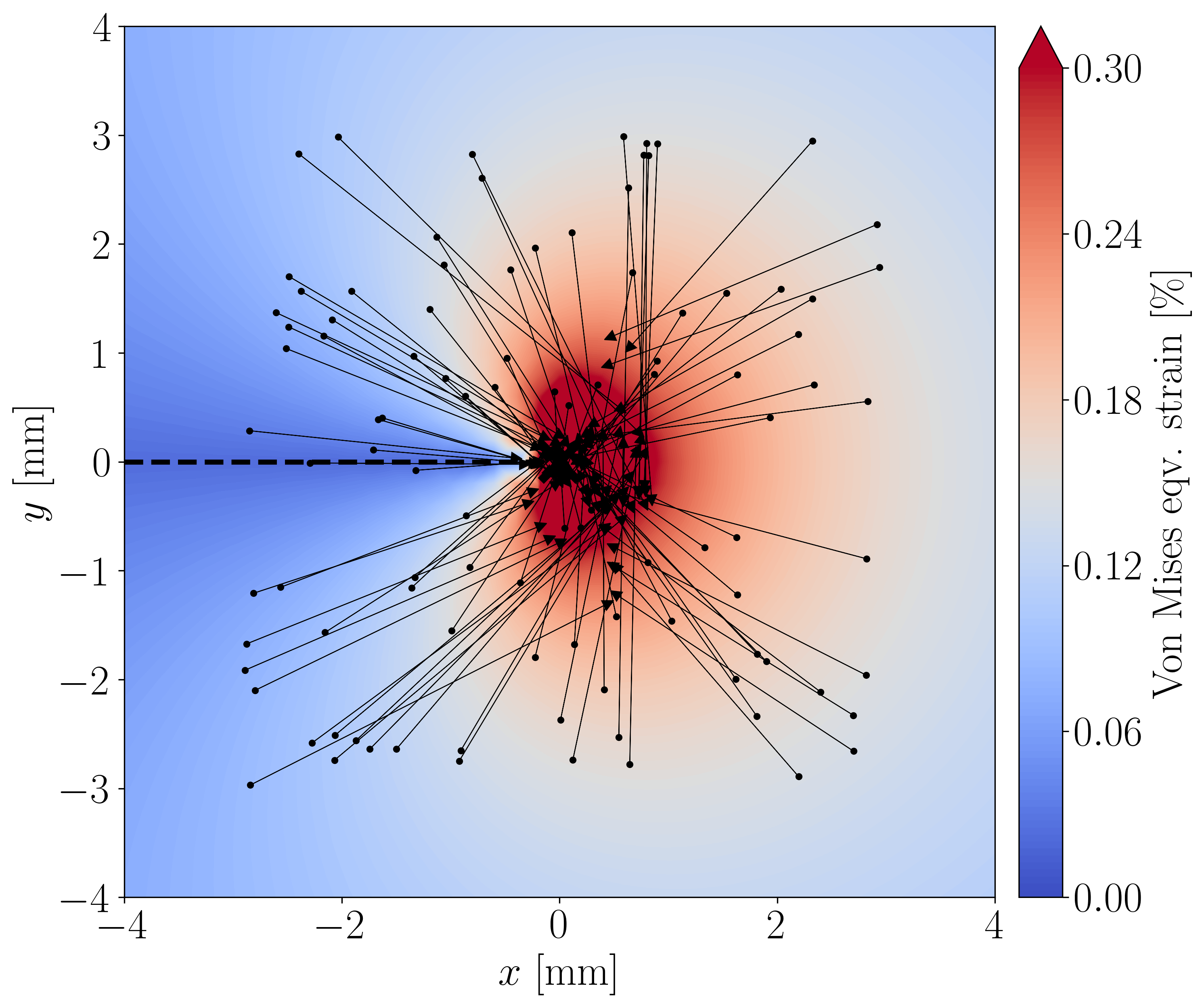}
    \hfill
    \includegraphics[width=0.49\textwidth]{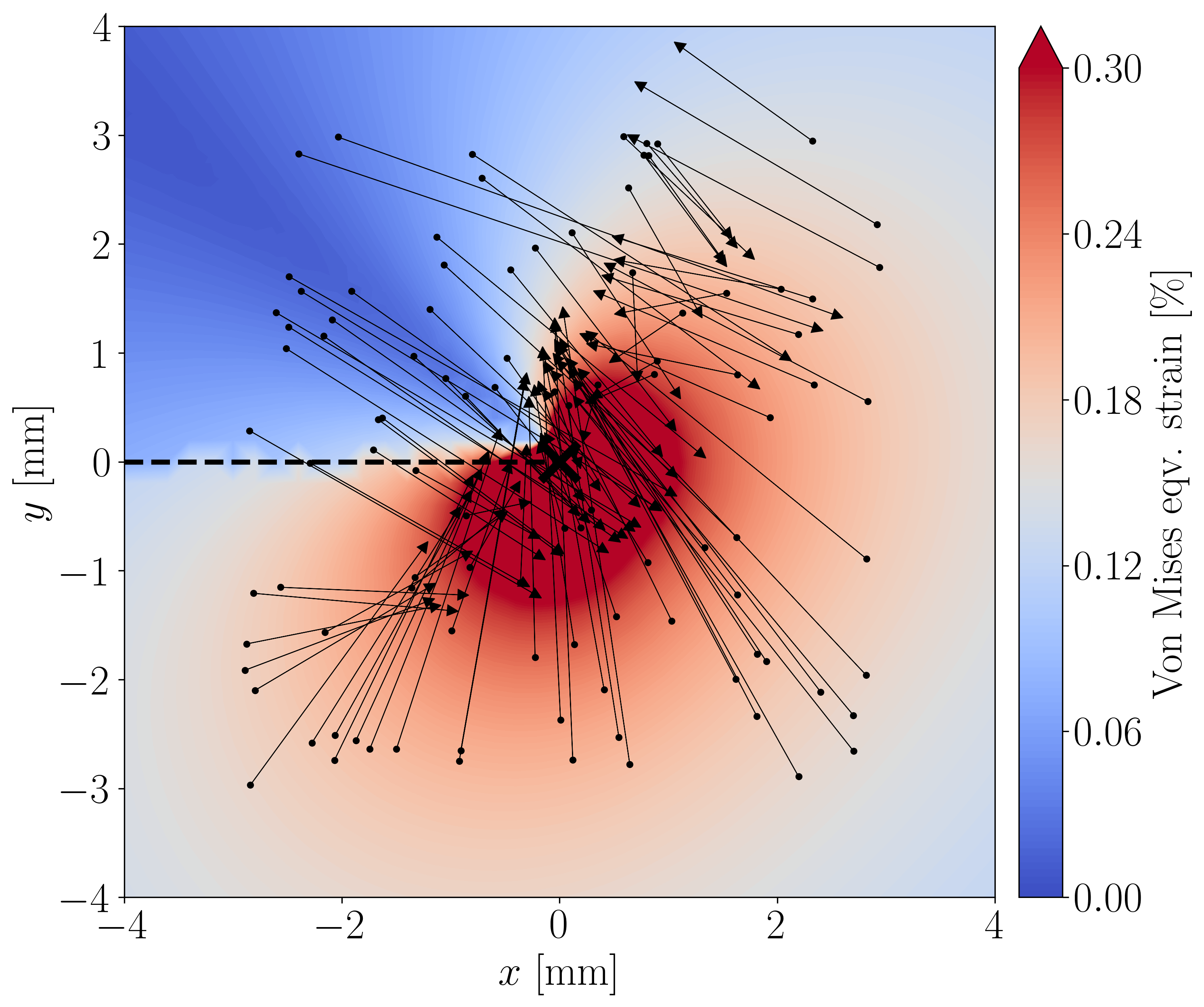}
    \caption{Correction vectors using the mode I formula \eqref{eq:symreg_mode_I} with $c_x^{\rm I} = 2$ and $c_y^{\rm I} = 3/2$ for $\sigma_{xx} = \sigma_{yy} = 10, \sigma_{xy} = 0$ [MPa] (left) and $\sigma_{xx} = \sigma_{yy} = \sigma_{xy} = 10$ [MPa] (right).}
    \label{fig:appendix_mode_I_vectors}
\end{figure}

\begin{figure}[htbp]
    \centering
    \includegraphics[width=0.49\textwidth]{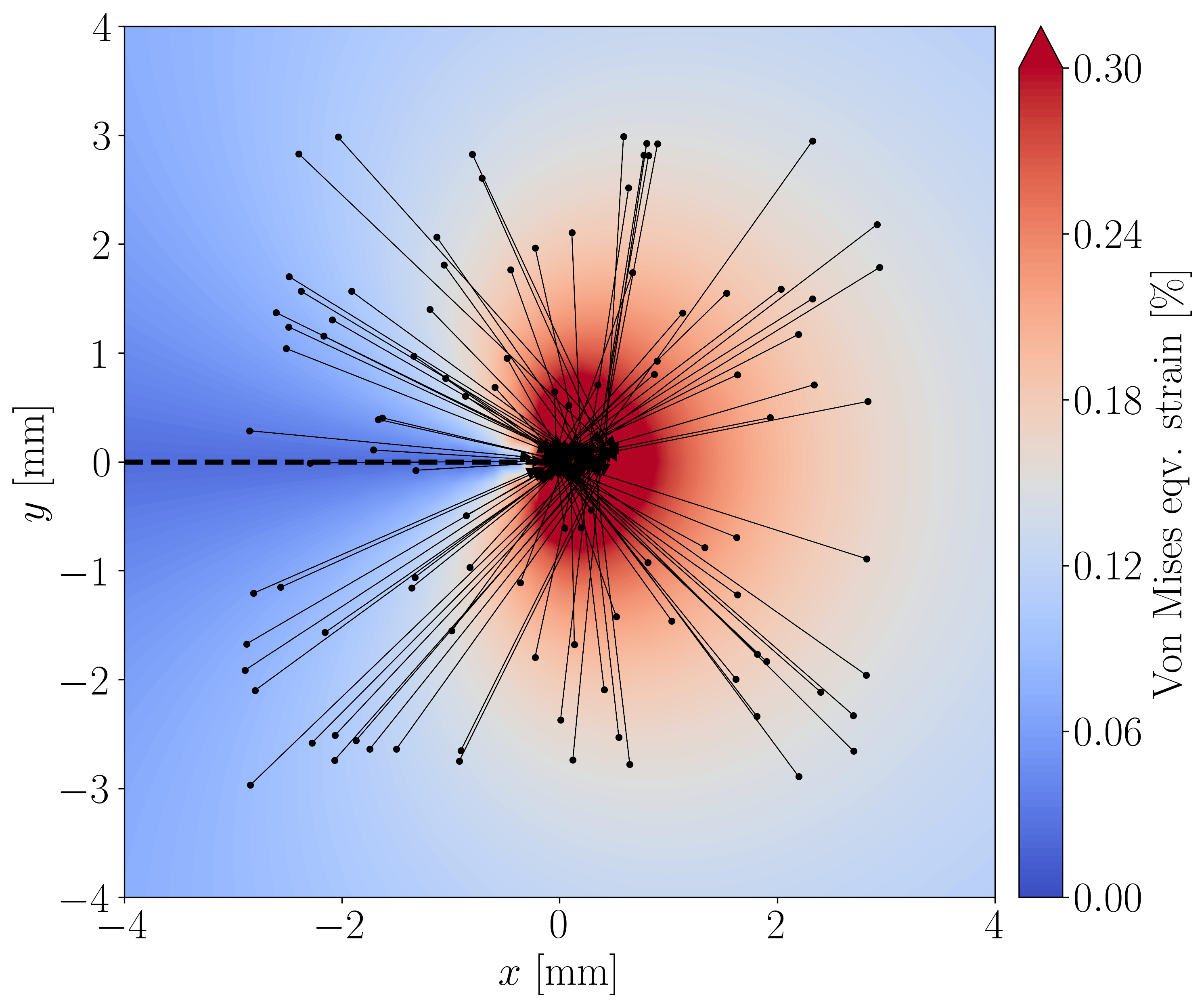}
    \hfill
    \includegraphics[width=0.49\textwidth]{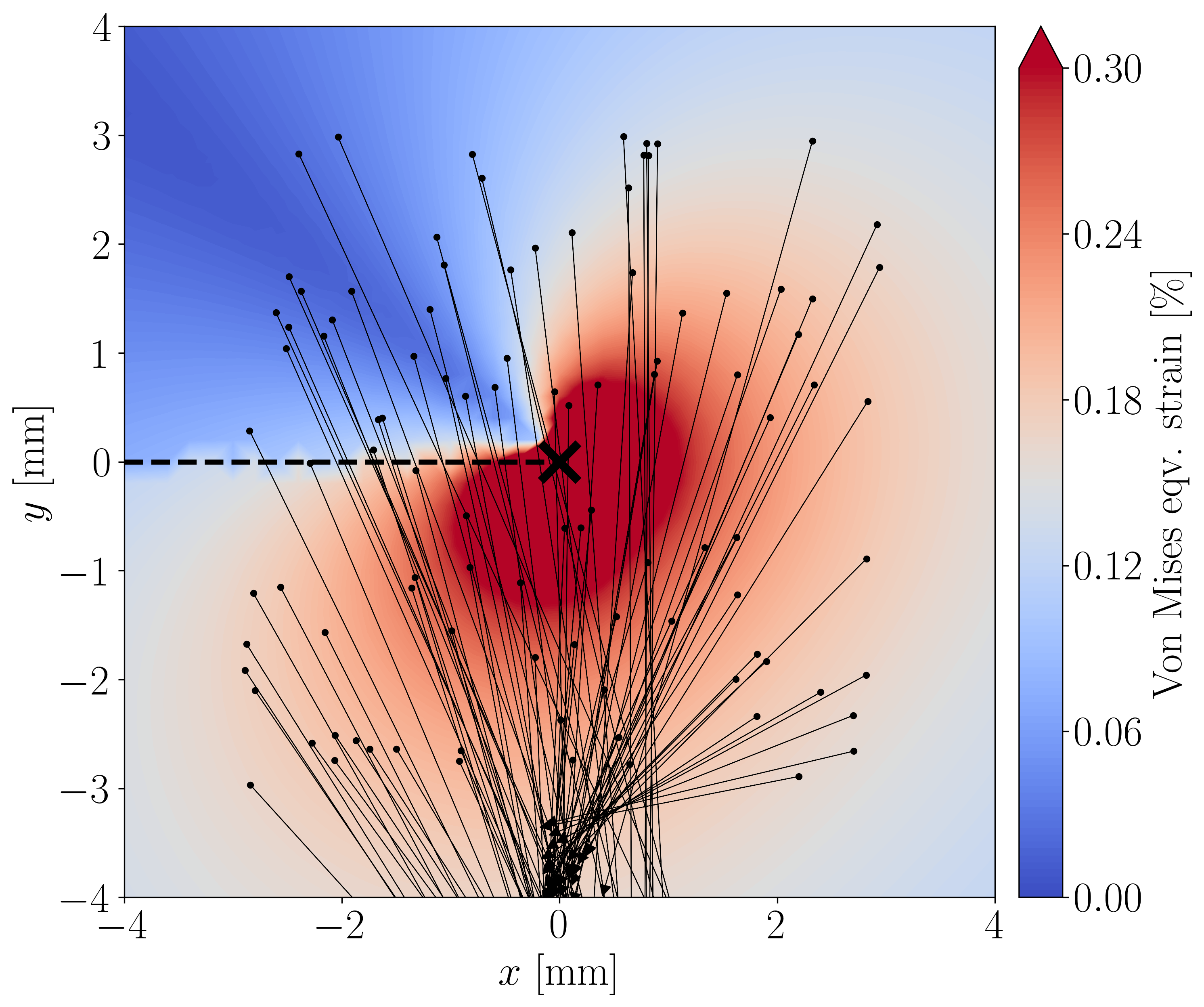}
    \caption{Correction vectors using the mode I formulas \#12 and \#6 for $d_x$ and $d_y$, respectively, for $\sigma_{xx} = \sigma_{yy} = 10, \sigma_{xy} = 0$ [MPa] (left) and $\sigma_{xx} = \sigma_{yy} = \sigma_{xy} = 10$ [MPa] (right).}
    \label{fig:appendix_mode_I_complex_vectors_fail}
\end{figure}

\begin{figure}[htbp]
    \centering
    \includegraphics[width=0.49\textwidth]{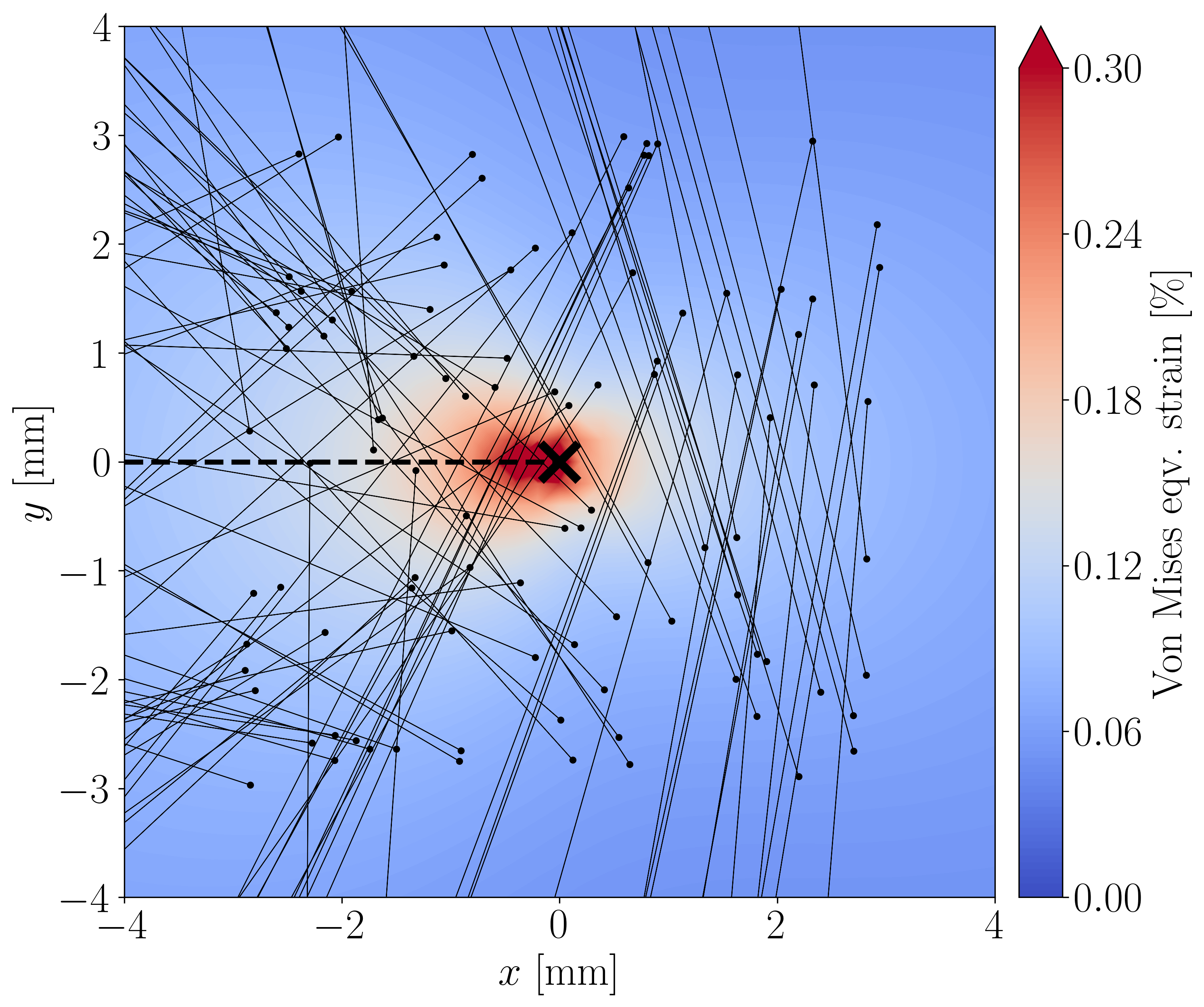}
    \caption{Correction vectors using the mode I Formula \eqref{eq:symreg_mode_I} with $c_x^{\rm I} = 1$ and $c_y^{\rm I} = 1$ for $\sigma_{xx} = \sigma_{yy} = 0, \sigma_{xy} = 10$ [MPa].}
    \label{fig:appendix_mode_I_vectors_fail_mode_II}
\end{figure}

In Section \ref{sec:results_vector_fields}, we only discussed the correction vector fields for selected mode I, II, and mixed-mode formulas. For example, we used Formula \ref{eq:symreg_mode_I} with constants equal to 1 for the mode I correction. With the optimized constants $c_x^{\rm I}=2, c_y^{\rm I}=3/2$, iterative correction is theoretically faster (see vector fields in Figure \ref{fig:appendix_mode_I_vectors} but less stable for experimental DIC data. When choosing more complex formulas such as \#12 for $d_x$ and \#6 for $d_y$ in Table \ref{table:appendix_eqs_pareto_I}, correction only works for pure mode I and not for mixed-mode anymore (see Figure \ref{fig:appendix_mode_I_complex_vectors_fail}. As mentioned in the present work, Formula \ref{eq:symreg_mode_I} works for all mode-I-dominated mixed-mode load cases but not for pure mode II (see Figure \ref{fig:appendix_mode_I_vectors_fail_mode_II}.

\newpage
\subsection*{Iterative correction for experimental full-field DIC data}
In Section \ref{sec:application_mDIC}, we showed convergence for two examples of high-resultion DIC data from uniaxial and biaxial fatigue crack growth experiments obtained from a robot carrying a DIC microscope. With less effort (but also less precision), our iterative crack tip correction method can also be applied to full-field data from the global 3D DIC system. Here, we show 4 snapshots at different stages of crack growth for the uniaxial experiment.

\begin{figure}[!htb]
    \centering
    \includegraphics[width=0.49\textwidth]{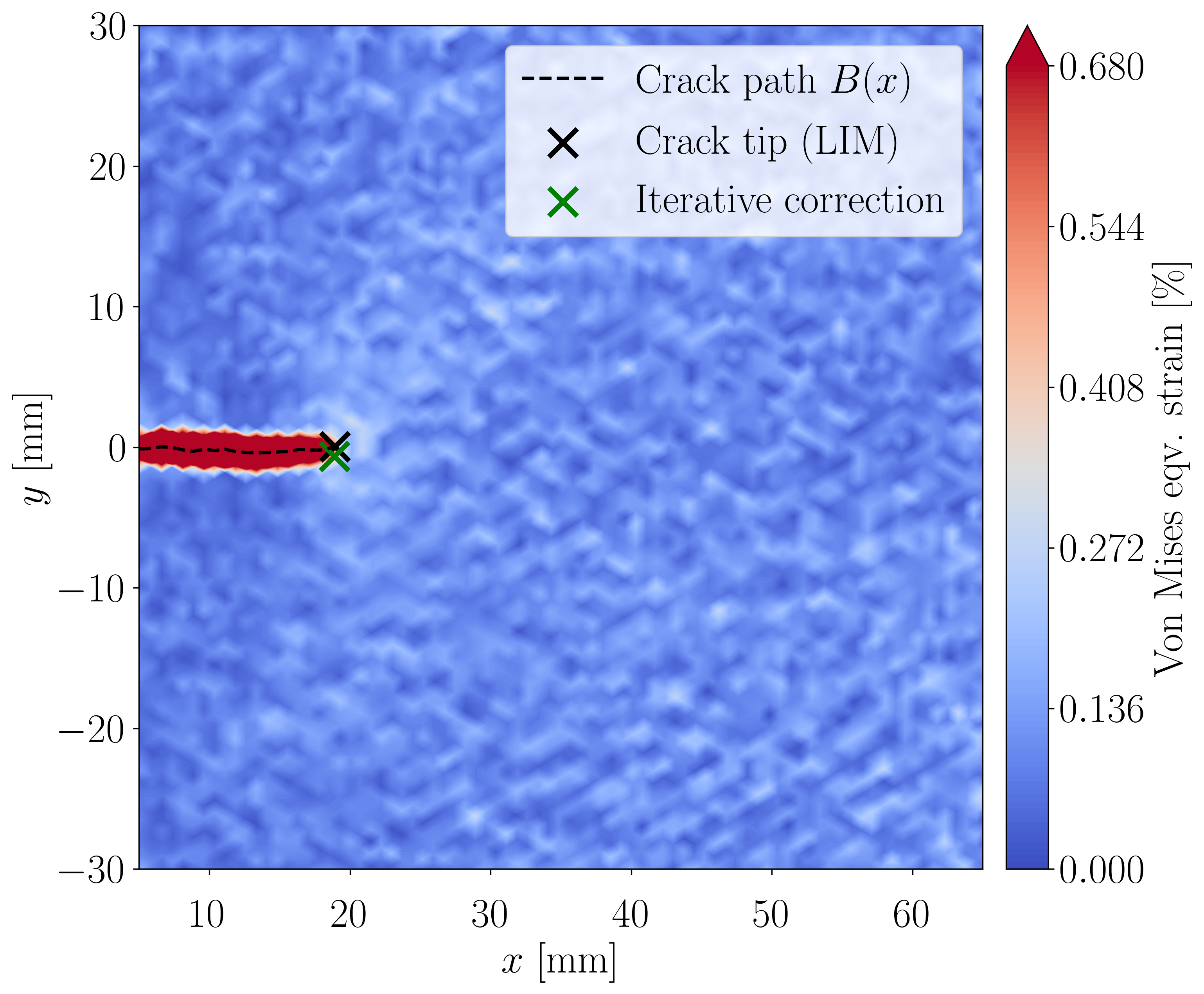}
    \includegraphics[width=0.49\textwidth]{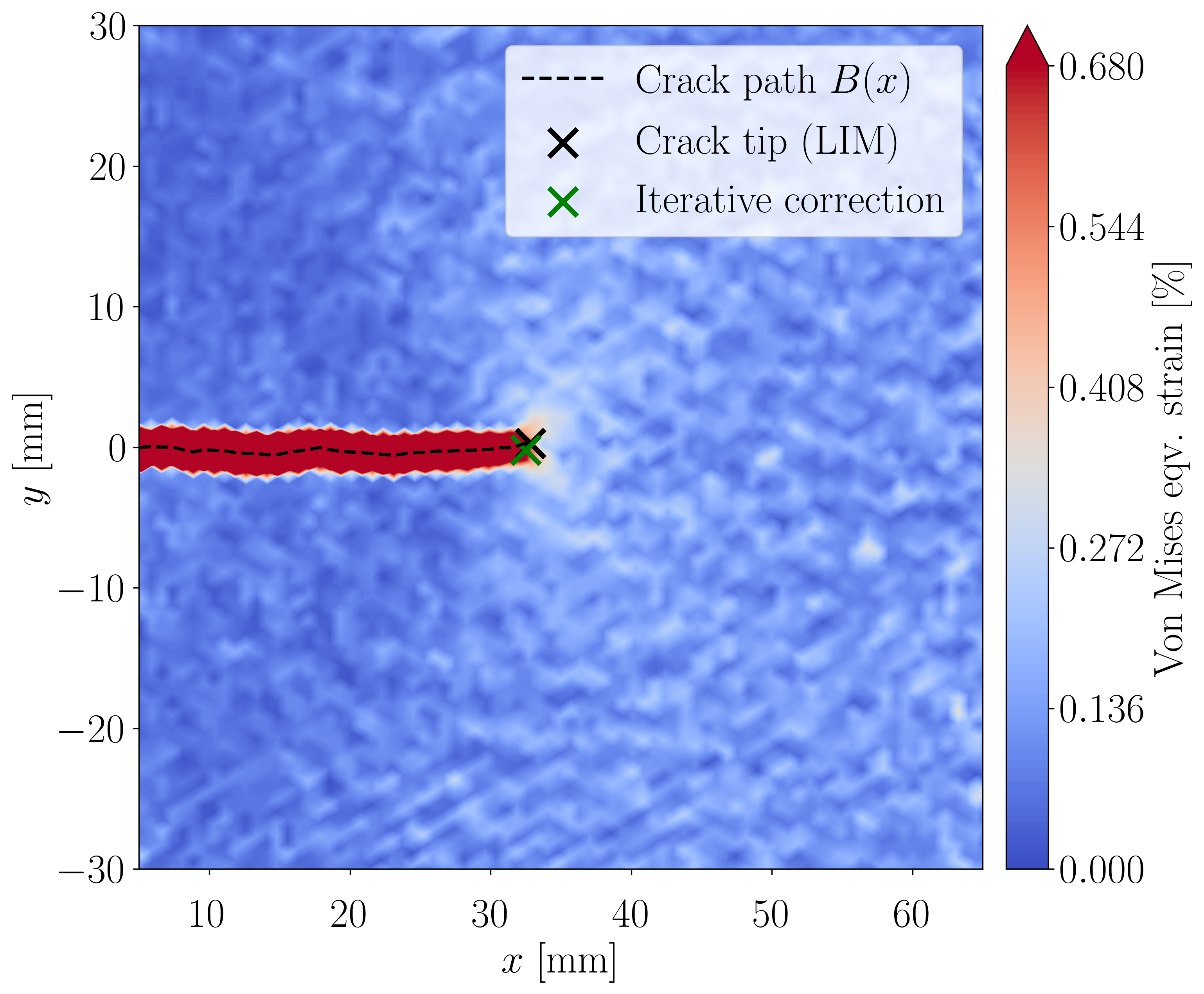}
    \includegraphics[width=0.49\textwidth]{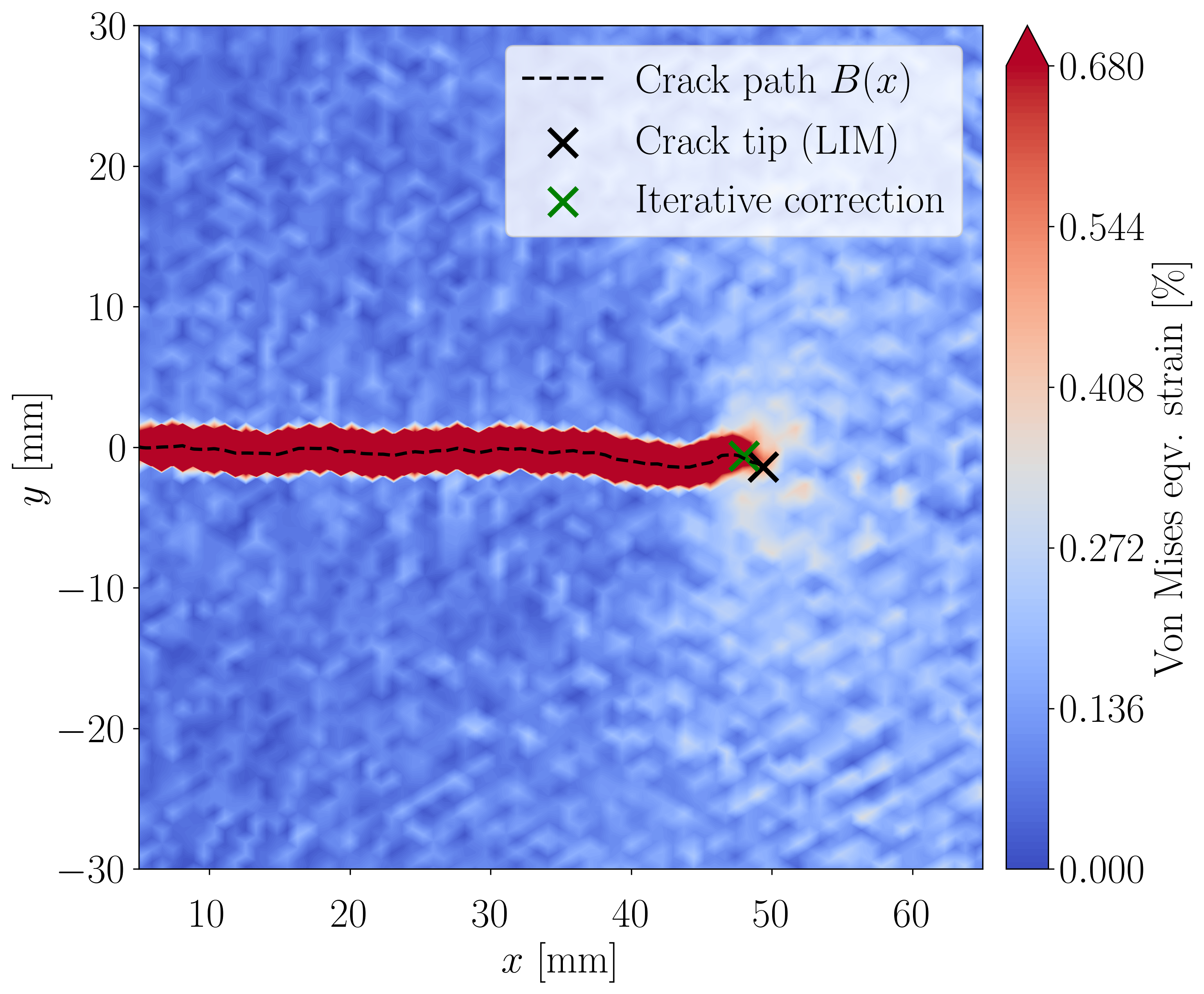}
    \includegraphics[width=0.49\textwidth]{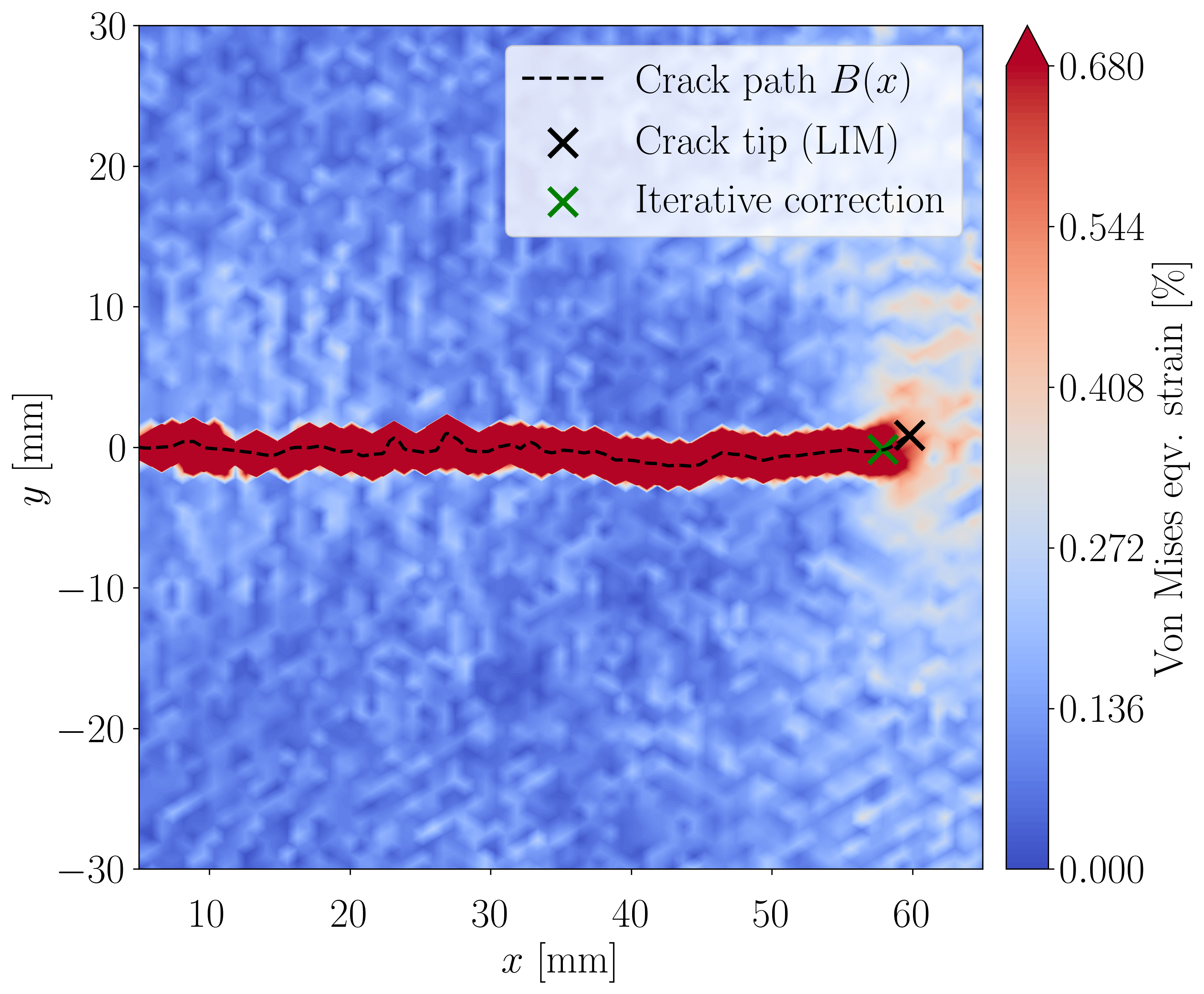}
    \caption{Iterative correction at different crack lenghts for full-field 3D-DIC data of uniaxial fatigue crack growth experiment using the mode I Formula \eqref{eq:symreg_mode_I} with $c_x^{\rm I} = c_y^{\rm I} = 1$.}
    \label{fig:appendix_3D_DIC}
\end{figure}

\newpage

\subsection*{More complex formulas}
\begin{figure}[htbp]
    \centering
    \begin{subfigure}[t]{0.42\textwidth}
        \includegraphics[width=0.95\textwidth]{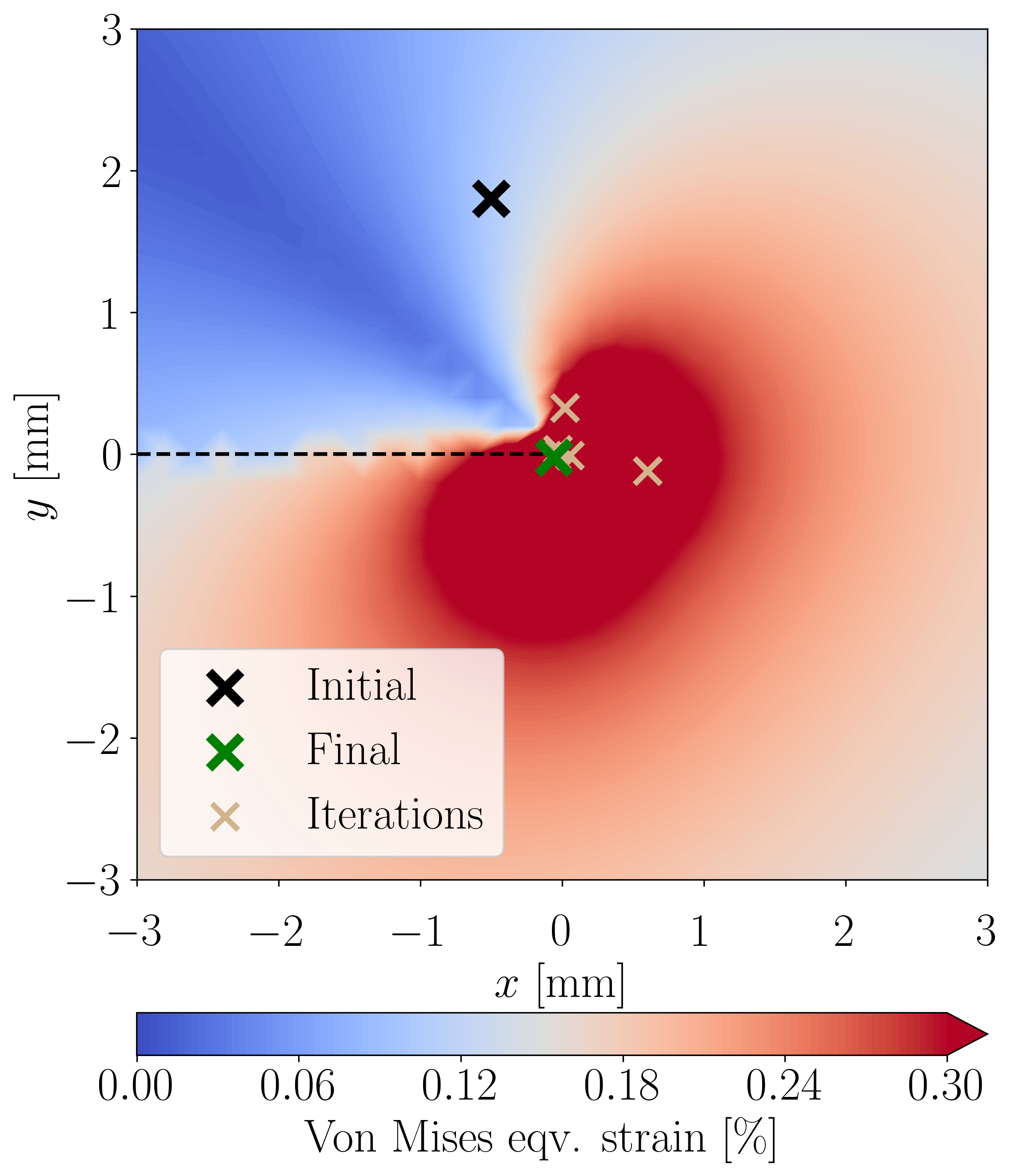}
        \caption{FE iterations}
    \end{subfigure}
    \hfill
    \begin{subfigure}[t]{0.49\textwidth}
        \includegraphics[width=0.95\textwidth]{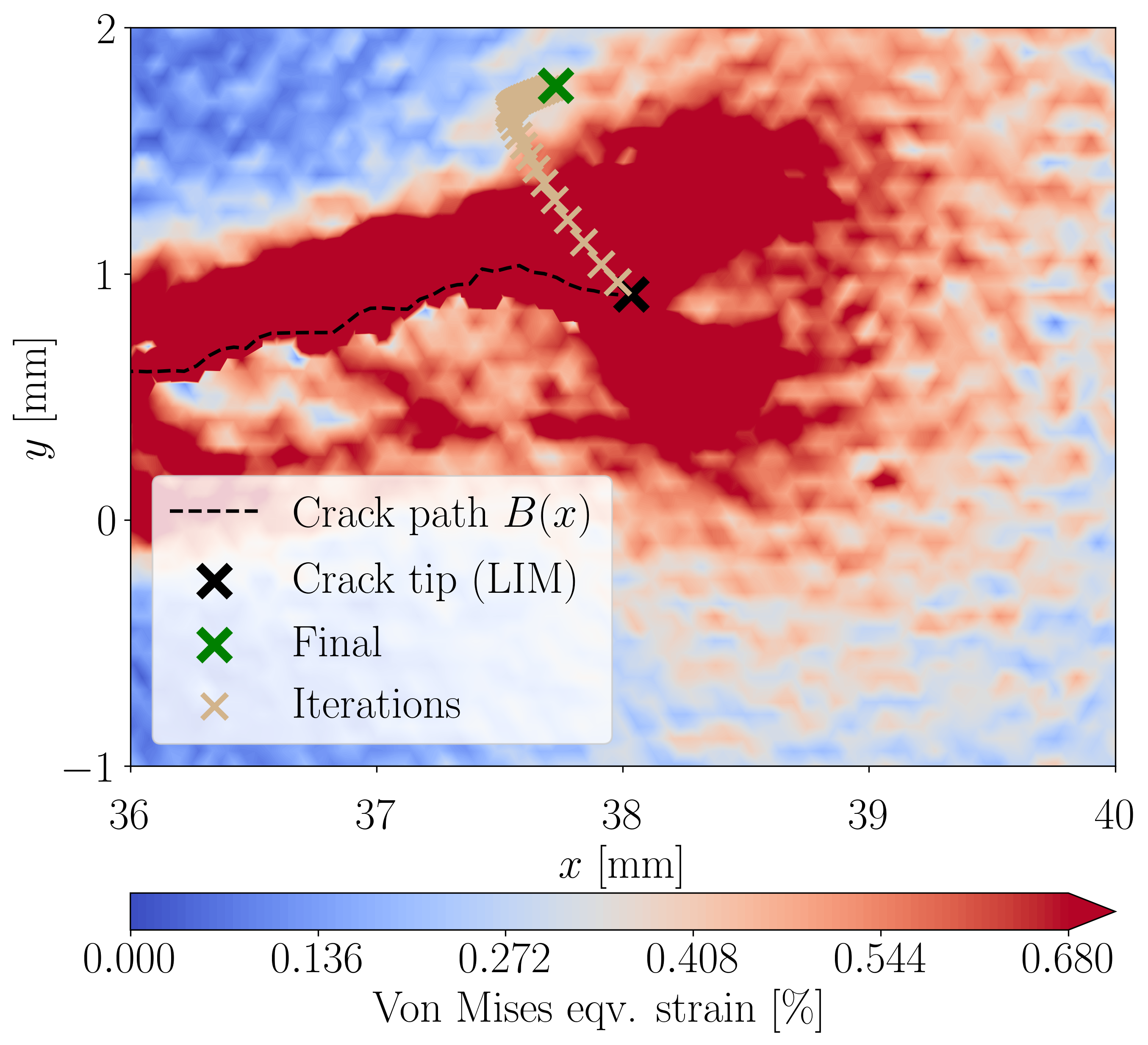}
        \caption{DIC iterations}
    \end{subfigure}
    \begin{subfigure}[t]{0.49\textwidth}
        \includegraphics[width=0.95\textwidth]{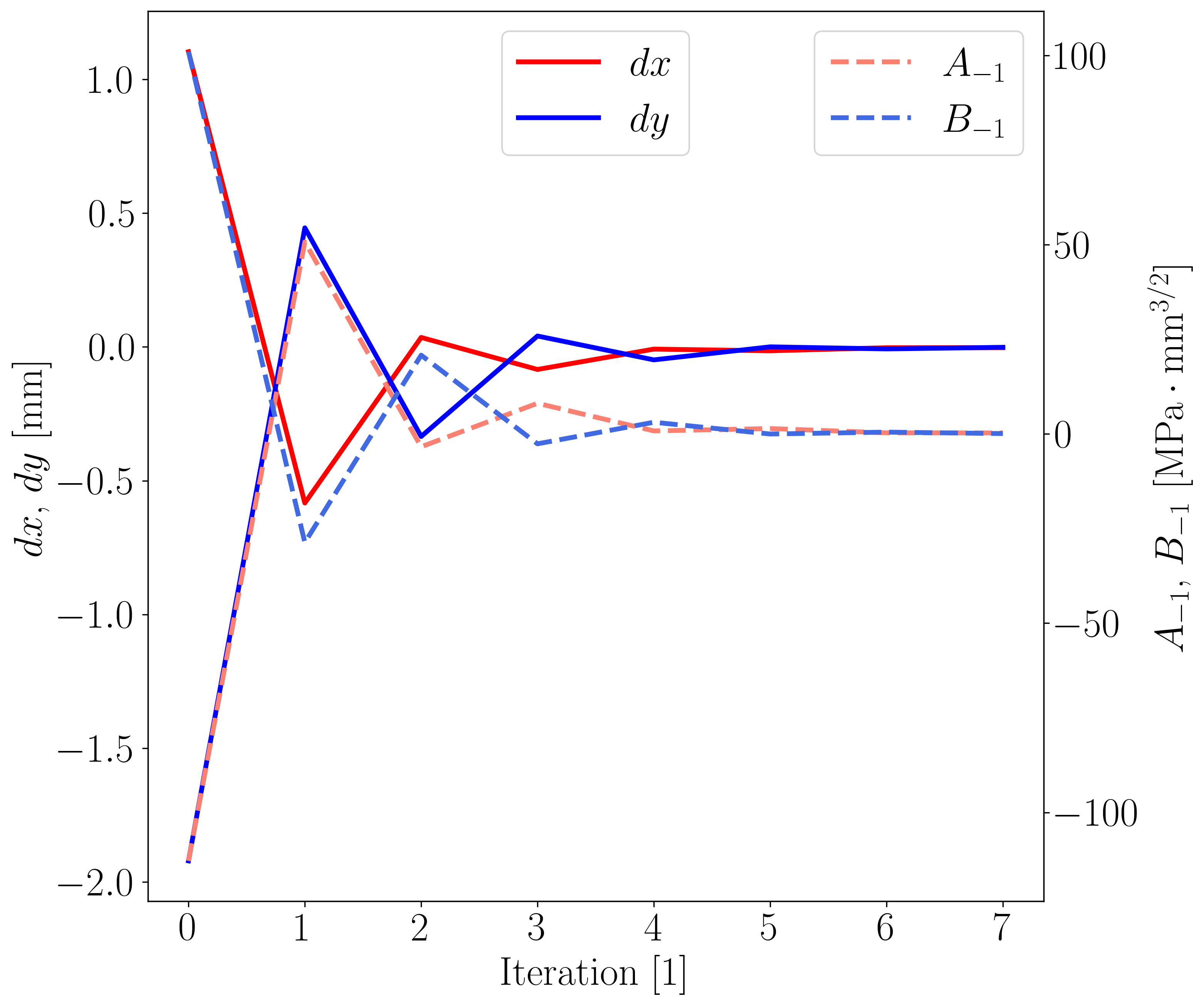}
        \caption{FE convergence}
    \end{subfigure}
    \hfill
    \begin{subfigure}[t]{0.49\textwidth}
        \includegraphics[width=0.95\textwidth]{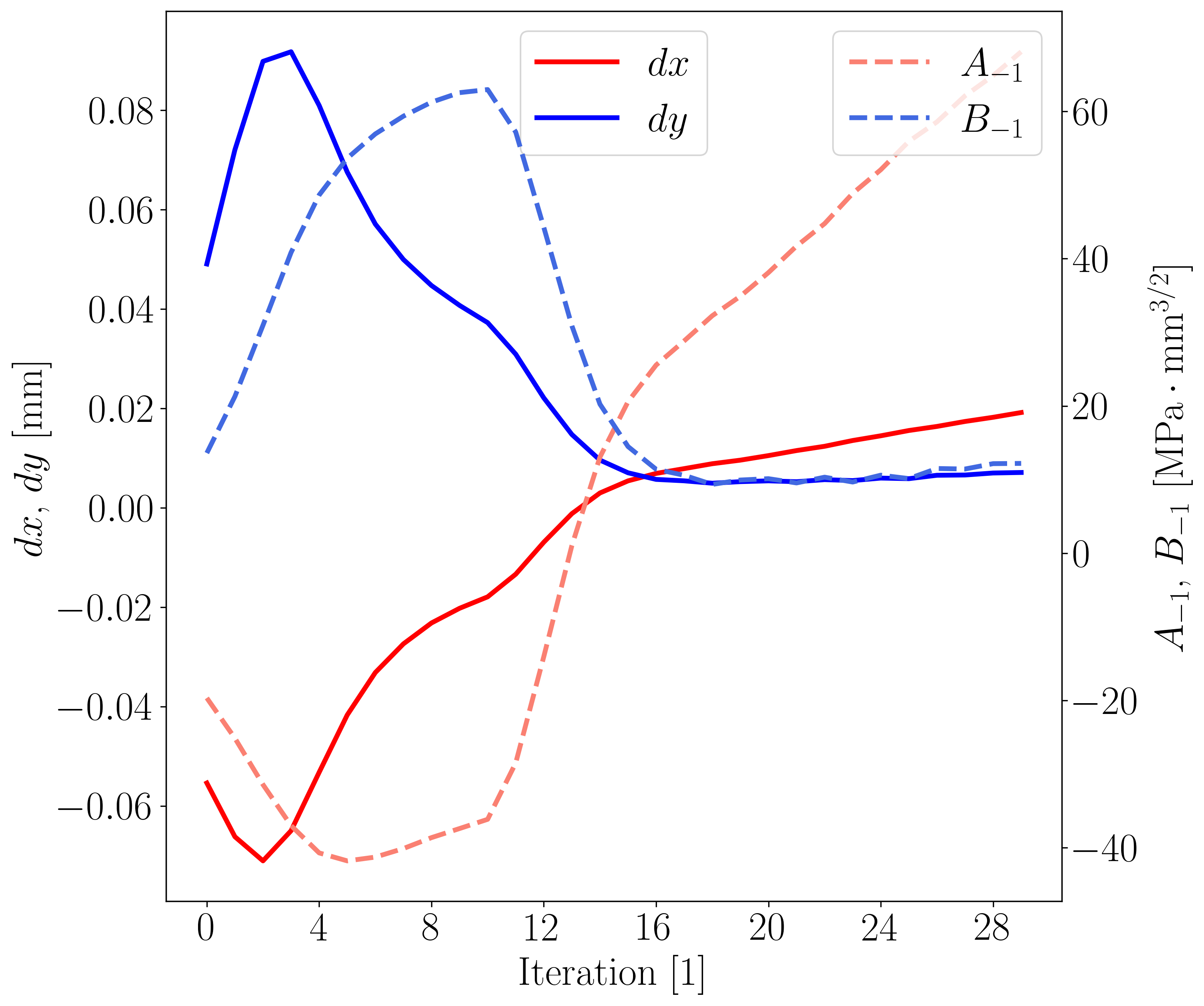}
        \caption{DIC non-convergence}
    \end{subfigure}
    \caption{Iterative correction using the complex mixed-mode formulas (\#11, \#8) for FE data of mixed-mode load case vs. DIC data of uniaxial test. Top: von Mises eqv. strain with crack tip correction iterations. Bottom: (Non-)convergence of $d_x, d_y$ and $A_{-1}, B_{-1}$ to zero. Left: FE. Right: DIC.}
    \label{fig:appendix_conv_more_complex}
\end{figure}

To demonstrate that the more complex formulas with a small RMSE often do not work equally well for experimental DIC data, we focus on the mixed-mode formula \#11 and \#8 for $d_x$ and $d_y$, respectively. Figure \ref{fig:appendix_conv_more_complex} shows that the correction works for FE data, but fails for DIC.

\end{document}